\documentclass[aps,prd,twocolumn,preprintnumbers,
superscriptaddress,showpacs,floatfix]{revtex4}

\usepackage{bm}
\usepackage{epsfig}
\usepackage{graphics}
\usepackage{amsmath,amssymb}
\usepackage{array}

\newcommand{\pdf}{\mathcal{P}}
\newcommand{\data}{{\bf D}}
\newcommand{\mdl}{\mathcal{M}}
\newcommand{\lsim}{\,\raise 0.4ex\hbox{$<$}
\kern -0.8em\lower 0.62ex\hbox{$\sim$}\,}
\newcommand{\params}{\boldsymbol{\theta}}

\newcommand{\like}{L}

\newcommand{\dr}{\textrm{d}}

\newcommand{\be}{\begin{equation}}
\newcommand{\ee}{\end{equation}}

\newcommand{\om}{\omega}

\newcommand{\ad}{\text{ad}}

\newcommand{\uf}{\;[10^{-5}]}
\newcommand{\ut}{\;[10^{-10}]}

\begin{document}

\preprint{UT-STPD-3/04}

\title{
Constraints on a mixed inflaton and curvaton
scenario \\ for the generation of the
curvature perturbation}

\author{George Lazarides}
\email{lazaride@eng.auth.gr}
\affiliation{Physics Division, School of
Technology, Aristotle University of
Thessaloniki, Thessaloniki 54124, Greece}
\author{Roberto Ruiz de Austri}
\email{r.ruizdeaustri@sheffield.ac.uk}
\affiliation{University of Sheffield, Department
of Physics and Astronomy, Hicks Building,
Hounsfield Road, Sheffield S3 7RH, England}
\author{Roberto Trotta}
\email{roberto.trotta@physics.unige.ch}
\affiliation{D\'epartement de Physique
Th\'eorique, Universit\'e de Gen\`eve, 24
quai Ernest Ansermet, 1211 Gen\`eve 4,
Switzerland}

\date{\today}

\begin{abstract}
We consider a simple supersymmetric grand
unified model which naturally solves the
strong CP and $\mu$ problems via a Peccei-Quinn
symmetry and leads to the standard realization
of hybrid inflation. We show that the
Peccei-Quinn field of this model can act as a
curvaton. In contrast to the standard curvaton
hypothesis, both the inflaton and the curvaton
contribute to the total curvature perturbation.
The model predicts the existence of an
isocurvature perturbation too which has mixed
correlation with the adiabatic one. The cold
dark matter of the universe is mostly
constituted by axions, which are produced at
the QCD phase transition, plus a small amount
of lightest sparticles. The predictions of the
model are confronted with the first-year
Wilkinson microwave anisotropy probe and other
cosmic microwave background radiation data. We
analyze in detail two representative choices of
parameters for our model and derive bounds on
the curvaton contribution to the adiabatic
perturbation. We find that, for the choice
which provides the best fitting of the data,
the curvaton contribution to the amplitude of
the adiabatic perturbation must be smaller than
about $67\%$ and the amplitude of the partial
curvature perturbation from the curvaton
smaller than $43.2\times 10^{-5}$ (both at
$95\%$ confidence level). The best-fit power
spectra are dominated by the adiabatic part of
the inflaton contribution. We use Bayesian
model comparison to show that this choice of
parameters is disfavored with respect to the
pure inflaton scale-invariant case with odds of
about 50 to 1. For the second choice of
parameters examined, the adiabatic mode is
dominated by the curvaton, but this choice is
strongly disfavored relative to the pure
inflaton scale-invariant case (with odds
of about $10^7$ to 1). We conclude that in the
present framework the perturbations must be
dominated by the adiabatic component from the
inflaton.

\end{abstract}

\pacs{12.10.-g,12.60.Jv,98.80.Cq}

\maketitle

\section{Introduction}
\label{intro}

\par
Inflation, which was originally proposed
\cite{guth} as a solution to the outstanding
problems of standard big bang cosmology and the
problem of unwanted relics, is in good agreement
with the recent measurements \cite{wmap1,wmap2}
on the angular power spectrum of the cosmic
microwave background radiation (CMBR). Moreover,
inflation is now established as the most likely
origin of the primordial density perturbation
from which structure formation in the universe
proceeded \cite{llbook}. According to the
usual assumption \cite{llbook,lectures}, this
perturbation is generated solely by the slowly
rolling inflaton field of the usual one-field
inflationary models and, thus, is expected to
be purely adiabatic. However, although adiabatic
perturbations are perfectly consistent with the
present data, the presence of a significant
isocurvature density perturbation cannot be
excluded \cite{wmap2,iso1}. In one-field
inflation, the perturbations are almost
Gaussian, in agreement with the current upper
bounds on non-Gaussianity from the CMBR data,
which though cannot exclude the presence of
appreciable non-Gaussianity (for a review, see
e.g. Ref.~\cite{nongauss}).

\par
Lately, the alternative possibility
\cite{curv1,curv2} that the adiabatic density
perturbations originate from the inflationary
perturbations of some light ``curvaton'' field
different from the inflaton has attracted much
attention. The curvaton density perturbations
can lead \cite{curv2,curv3}, after curvaton
decay, to isocurvature perturbations in the
densities of the various components of the
cosmic fluid. In the simplest case, the
residual isocurvature perturbations are either
fully correlated or fully anti-correlated with
the adiabatic density perturbation. In general
models, however, the correlation can be mixed
(see e.g. Ref.~\cite{dllr1}). In the curvaton
scenario, significant non-Gaussianity may also
appear. The main reason for advocating the
curvaton hypothesis is that it makes
\cite{dimo} the task of constructing viable
models of inflation much easier, since it
liberates us from the very restrictive
requirement that the inflaton is responsible
for the curvature perturbations.

\par
In a recent paper \cite{dllr1}, a simple
extension \cite{rsym} of the minimal
supersymmetric standard model (MSSM) which
naturally and simultaneously solves the strong
CP and $\mu$ problems via a Peccei-Quinn (PQ)
\cite{pq} and a continuous R symmetry was
considered within the general framework of the
standard supersymmetric (SUSY) version
\cite{lyth,dss} of hybrid inflation
\cite{hybrid}. It was shown that, in this model,
the PQ field, which breaks spontaneously the
PQ symmetry, can successfully act as a curvaton
generating the total curvature perturbation
in the universe in accordance with the cosmic
background explorer (COBE) measurements
\cite{cobe}. The (intermediate) PQ scale
is generated by invoking supergravity (SUGRA)
and the PQ field corresponds to a flat
direction in field space lifted by
non-renormalizable interactions. Moreover,
the $\mu$ parameter of MSSM is generated
\cite{kn} from the PQ scale.

\par
We feel that the standard curvaton hypothesis
\cite{curv1,curv2}, which insists that the
total curvature perturbation originates
solely from the curvaton, can also be quite
restrictive and not so natural. Indeed, in
accordance to this hypothesis, one needs to
impose \cite{dimo} an upper bound on the
inflationary scale in order to ensure that the
perturbation from the inflaton is negligible.
This bound can be quite strong especially if
the slow-roll parameter $\epsilon$ (see e.g.
Ref.~\cite{lectures}) happens to be very small,
which holds in many cases. In generic models,
one would expect that all the scalar fields
which are essentially massless during inflation
contribute to the total curvature perturbation.
So, in the presence a PQ field, which can be
kept \cite{dllr1} light during the relevant
part of inflation, it is natural to assume that
the adiabatic density perturbation is partly
due to this field and partly to the inflaton.

\par
There is yet another reason for abandoning the
strict curvaton hypothesis. The recent
measurements on the CMBR by the Wilkinson
microwave anisotropy probe (WMAP) satellite
\cite{wmap1} have considerably strengthened
\cite{wmap2,iso1,gm} the bound on the
isocurvature perturbation which was obtained
\cite{iso2,iso3} by using the pre-WMAP CMBR
data. As a consequence, the viability
of many curvaton models is in doubt. However,
allowing a significant part of the total
curvature perturbation in the universe to
originate from the inflaton, we can hopefully
relax the tension between these models and
the present WMAP data without losing the main
advantage of the curvaton hypothesis, which
is that it facilitate the construction of
viable inflationary models (see
Ref.~\cite{mixed} for recent investigations of
this possibility).

\par
The PQ curvaton model of Ref.~\cite{dllr1}
predicts an isocurvature perturbation of
mixed correlation with the curvature
perturbation. The extended set of
pre-WMAP CMBR and other data which was used
\cite{dllr1} to restrict the isocurvature
perturbation in the model of
Ref.~\cite{dllr1} led to the exclusion of a
major part of the available parameter space.
Including the more restrictive recent WMAP
measurements, the allowed parameter space
will certainly be further reduced. It is,
moreover, quite possible that the model is
even totally excluded by these new data. So,
the departure from the strict curvaton
hypothesis may prove to be vital for this
particular curvaton scheme.

\par
In this paper, we will extend the PQ curvaton
model of Ref.~\cite{dllr1} by embedding it
into the concrete SUSY grand unified theory
(GUT) model studied in Ref.~\cite{hier}, which
is based on the left-right (LR) symmetric gauge
group $G_{LR}={\rm SU}(3)_c\times {\rm SU}(2)_L
\times{\rm SU}(2)_R\times {\rm U}(1)_{B-L}$.
This model leads \cite{hier} naturally to
standard SUSY hybrid inflation \cite{lyth,dss}.
After the end of inflation, the inflaton
performs damped oscillations about the SUSY
vacuum and eventually decays into right handed
neutrino superfields reheating the universe.
The subsequent decay of these superfields to
a lepton and an electroweak Higgs superfield
generates \cite{leptoinf} a primordial lepton
asymmetry \cite{lepto} which is then partly
converted into baryon asymmetry by
non-perturbative electroweak sphaleron effects.
The observed baryon asymmetry of the universe
(BAU) can then be easily reproduced \cite{hier}
in accord with the data on neutrino masses and
mixing. At reheating, gravitinos are also
produced thermally. They decay in the late
universe leading to lightest sparticles (LSPs),
which contribute to the cold dark matter (CDM)
in the universe. For simplicity, we assume
that this is the only source of relic LSPs
neglecting their direct thermal production.
Due to the presence of the PQ symmetry, our
model contains axions which come into play at
the QCD phase transition and can also
contribute to CDM.

\par
The PQ field of our model can acquire a
super-horizon spectrum of perturbations from
inflation provided that it is effectively
massless during the relevant part of inflation.
It can thus act as curvaton contributing to
the total curvature perturbation together with
the inflaton. We study the evolution of the PQ
field during and after inflation by including
corrections \cite{lyth,crisis,drt95} to the PQ
potential which originate from the SUSY
breaking in the early universe caused by the
presence of a finite energy density. We assume
that these corrections are (somewhat)
suppressed, which is \cite{noscale} indeed the
case if specific K\"{a}hler potentials are
used.

\par
The requirement that the PQ field is
essentially massless during inflation yields
\cite{dllr1}, for given values of the other
parameters, an upper bound on the possible
values of this field at the end of inflation.
Moreover, it implies that, as inflation
terminates, the PQ field emerges \cite{dllr1}
with negligible velocity. There is also a
lower bound on the value of the PQ field at
the end of inflation below which the classical
equation of motion during inflation for the
mean value of this field in a region of fixed
size somewhat bigger than the size of the de
Sitter horizon ceases \cite{dllr1} to be valid.
This is due to the fact that the quantum
perturbations of the PQ field from inflation
overshadow its classical kinetic energy
density. We will exclude this quantum regime
since the calculation of the spectral index of
the partial curvature perturbation from the
curvaton field in this regime is not so clear.

\par
The values of the PQ field at the end of
inflation which are not excluded by the above
considerations can be classified according
to whether they lead to the PQ vacua or the
trivial (false) vacuum. We find that,
generically, there exist two separate bands
of such values leading to the correct (PQ)
vacua. One of them lies at relatively low
values of the curvaton field at the end of
inflation, while the other lies at values
which are considerably higher. In all other
cases, the system ends up in the wrong
(trivial) vacuum and thus the corresponding
values of the PQ field at the end of
inflation must be excluded. Our numerical
findings show that the (approximate) COBE
constraint on the CMBR can be satisfied
only within the upper allowed band. This
constraint receives contributions not only
from the curvature perturbation but also
from the isocurvature one and the cross
correlation of the two. Note, however, that
this constraint is quite approximate and
can be considered only as an indicative
criterion.

\par
The amplitude and spectral index of the
partial curvature perturbation from the
inflaton are calculated by employing the
analysis of Ref.~\cite{lectures} slightly
modified to allow for the possibility that
the slow-roll conditions are violated and,
thus, hybrid inflation ends before
reaching the instability point on the
inflationary trajectory. The partial
curvature perturbation from the curvaton
is treated in a more accurate way than in
Ref.~\cite{dllr1}. In particular, the
evolution during inflation of the
perturbation acquired by the PQ field from
inflation when our present horizon scale
crossed outside the inflationary horizon
is considered. It is described by the
classical equation of motion for this field
in the slow-roll approximation. Solving
this equation, we can find the perturbation
in the value of the curvaton field at the
end of inflation. This same calculation
yields the spectral index for the curvaton
too. For any given value of the
PQ field at the termination of inflation, we
take the perturbed value too and follow the
subsequent evolution of both these fields
until the time of the curvaton decay. This
yields the amplitude of the partial curvature
perturbation from the curvaton. The total
curvature perturbation is then given by the
appropriate weighted sum of these two
uncorrelated perturbations.

\par
As mentioned already, the baryons and the
LSPs in our model originate from reheating.
They thus inherit the partial curvature
perturbation of the oscillating and decaying
inflaton, which is different from the total
curvature perturbation due to the presence of
the curvaton. As a consequence, the baryons
and LSPs acquire an isocurvature perturbation
of mixed correlation with the total curvature
perturbation. The CDM in our model contains
also axions carrying an isocurvature
perturbation which is uncorrelated with the
perturbations from the inflaton and the
curvaton. The amplitude and spectral index of
this isocurvature perturbation is evaluated by
following the analysis of Ref.~\cite{dllr1}.
We see that, in our model, the correlation of
the total adiabatic and isocurvature
perturbations is mixed.

\par
For given values of all the other independent
parameters, we take a grid of values of
the curvaton field at the end of inflation and
the amplitude of partial curvature perturbation
from the inflaton which cover the upper or the
lower allowed band. We calculate the total CMBR
angular power spectrum for each point in this
grid. The predictions from each band are
confronted with the CMBR temperature (TT) and
temperature-polarization (TE) cross correlation
angular power spectra from the first-year WMAP
observations \cite{wmap1} as well as the CMBR
data on smaller scales from the arcminute
cosmology bolometer array receiver (ACBAR)
\cite{acbar1,acbar2} and the cosmic background
imager (CBI) \cite{cbi} experiments. We then
study the implications of the resulting
restrictions on the various parameters of the
model. We also employ Bayesian model testing to
compare our model with the standard pure
inflaton single-field inflationary model with
scale-invariant perturbations. We are
particularly interested to see whether the
data favor the presence of a non-vanishing
curvaton contribution to the adiabatic
perturbation.

\par
The paper is organized as follows. In
Sec.~\ref{sec:model}, we outline the salient
features of our LR SUSY GUT model which solves
the strong CP and $\mu$ problems via a PQ
symmetry and leads to the standard version of
SUSY hybrid inflation. The evolution of the PQ
field during and after inflation as well as its
final decay into light particles are sketched
in Sec.~\ref{sec:early}.
Section~\ref{sec:curvature} is devoted to the
evaluation of the total curvature perturbation
which, in our case, is partly due to the
inflaton and partly to the curvaton. In
Sec.~\ref{sec:isocurvature}, we estimate the
isocurvature perturbations in the relic
density of the baryons, the LSPs and the
axions. The total CMBR angular power spectrum
predicted by our model is discussed in
Sec.~\ref{sec:spectrum}. Our numerical
calculation and results are presented and
discussed in Sec.~\ref{sec:numerics}, and our
conclusions are summarized in
Sec.~\ref{sec:conclusion}. Finally, in the
Appendix, we review some useful concepts and
results from Bayesian statistics.

\section{The left-right SUSY GUT model}
\label{sec:model}

\par
We will adopt here the SUSY GUT model of
Ref.~\cite{hier} (see also Ref.~\cite{trieste})
which is based on the LR symmetric gauge group
$G_{LR}$. The ${\rm SU}(2)_{L}$ doublet left
handed quark and lepton superfields are denoted
by $q_i$ and $l_i$ respectively, whereas the
${\rm SU}(2)_{R}$ doublet antiquark and
antilepton superfields by $q_i^c$ and $l^c_i$
respectively ($i$=1,2,3 is the family index).
The electroweak Higgs superfield $h$ belongs to
a bidoublet $(2,2)_0$ representation of
${\rm SU}(2)_L\times {\rm SU}(2)_R\times
{\rm U}(1)_{B-L}$.

\par
The breaking of $G_{LR}$ to the standard model
(SM) gauge group $G_{\rm SM}$, at a superheavy
scale $M\sim 10^{16}~{\rm{GeV}}$, is achieved
through the superpotential
\begin{equation}
\delta W_1=\kappa S(l_H^c\bar l_H^{c}-M^2),
\label{W1}
\end{equation}
where $l_H^c$, $\bar l_H^{c}$ is a conjugate pair of
${\rm SU}(2)_R$ doublet left handed Higgs superfields
with $B-L$ charges equal to $1,-1$ respectively, and
$S$ is a gauge singlet left handed superfield. The
dimensionless coupling constant $\kappa$ and the
mass parameter $M$ can be made real and positive by
suitable rephasing of the fields. The SUSY minima of
the scalar potential lie on the D-flat direction
$l_H^c=\bar l_H^{c*}$ at $\langle S\rangle=0$,
$|\langle l_H^c\rangle|=|\langle\bar l_H^{c}\rangle|
=M$.

\par
The model also contains two extra gauge singlet left
handed superfields $N$ and $\bar{N}$ for solving
\cite{rsym} the $\mu$ problem of MSSM via a
PQ symmetry \cite{pq}, which also solves the strong
CP problem. They have the following superpotential
couplings:
\begin{equation}
\delta W_2=\frac{\lambda N^2\bar{N}^2}{2m_P}+
\frac{\beta N^2h^2}{2m_P},
\label{W2}
\end{equation}
where $\lambda$ and $\beta$ are dimensionless
parameters, which can be made real and positive
by an appropriate redefinition of the phases of
the superfields and $m_{\rm P}\simeq 2.44\times
10^{18}~{\rm GeV}$ is the reduced Planck mass.

\par
In addition to $G_{LR}$, the model possesses three
global ${\rm U}(1)$ symmetries, namely an anomalous
PQ symmetry ${\rm U}(1)_{\rm PQ}$, a non-anomalous
R symmetry ${\rm U}(1)_R$, and the baryon number
symmetry ${\rm U}(1)_B$. The PQ and R charges
of the various superfields are
\begin{eqnarray}
PQ:~q^c,l^c,S,l_H^c,\bar l_H^c(0),~h,\bar N(1),~q,l,
N(-1); \nonumber \\
R:~h,l_H^c,\bar l_H^c,\bar{N}(0),~q,q^c,l,l^c,N(1/2),
~S(1).
\label{charges}
\end{eqnarray}
Note that global continuous symmetries such as our
PQ and R symmetry can effectively arise
\cite{continuous} from the rich
discrete symmetry groups encountered in many
compactified string theories (see e.g.
Ref.~\cite{discrete}).

\par
It is well known that the superpotential in
Eq.~(\ref{W1}) leads \cite{lyth,dss} naturally
to the standard SUSY realization of hybrid
inflation \cite{hybrid}. In particular, the
scalar potential which is derived from it
possesses a built-in classically flat valley of
minima at $l_H^c=\bar l_H^c=0$ and for $|S|$
greater than a critical (instability) value
$S_c=M$. This valley can serve as inflationary
path. Indeed, the constant tree-level potential
energy density $\kappa^{2}M^{4}$ on this path
can cause exponential expansion of the
universe. Moreover, since this constant energy
density breaks SUSY, there are important
radiative corrections \cite{dss} which provide
a logarithmic slope along the inflationary
trajectory necessary for driving the system
towards the vacua.

\par
It should be noted that the SUSY GUT model
considered here does not predict the existence
of topological defects such as magnetic
monopoles or cosmic strings. In the opposite
case, these defects would have been copiously
produced \cite{smooth} at the end of hybrid
inflation. The overproduction of magnetic
monopoles, in particular, would have led to a
cosmological catastrophe and a modification
\cite{jean,smooth} of the standard realization
of SUSY hybrid inflation would be needed to
avoid this problem. This happens in higher
gauge groups such as the Pati-Salam group,
which predicts the existence of monopoles
carrying two units of Dirac magnetic charge
\cite{magg}. Cosmic strings, on the other hand,
which are generically present in many GUT
models \cite{kibble,generic}, would contribute
to the cosmological perturbations leading
\cite{mairi} to extra restrictions on the
parameters of the model. The reason that our
model does not predict cosmic strings is that
the $G_{LR}$ breaking is achieved by a
conjugate pair of ${\rm SU}(2)_R$ doublets
with $B-L=1,-1$ which also break the $Z_2$
subgroup of ${\rm U}(1)_{B-L}$. This $Z_2$,
which does not belong to $G_{\rm SM}$, would
have been left unbroken if, alternatively, we
had used a pair of ${\rm SU}(2)_R$ triplets
with $B-L=2,-2$ for this breaking. This would
have led to the presence of $Z_2$ cosmic
strings (compare with the $Z_2$ cosmic strings
encountered in the ${\rm SO}(10)$ GUT model of
Ref.~\cite{kibble}).

\par
The part of the tree-level scalar potential which
is relevant for the PQ symmetry breaking is derived
from the superpotential coupling
$\lambda N^2\bar{N}^2/2m_{\rm P}$ in Eq.~(\ref{W2})
and, after soft SUSY breaking mediated by minimal
supergravity (SUGRA), is given by \cite{rsym}
\begin{equation}
V_{\rm PQ}=\frac{1}{2}m_{3/2}^2\phi^2\left(
1-\frac{|A|\lambda\phi^2}{8m_{3/2}m_P}+
\frac{\lambda^2\phi^4}{16m_{3/2}^2m_P^2}\right),
\label{PQ-pot}
\end{equation}
where $m_{3/2}\sim 1~{\rm TeV}$ is the gravitino
mass and $A$ is the dimensionless coefficient of
the soft SUSY breaking term corresponding to the
superpotential term $\lambda N^2\bar{N}^2/
2m_{\rm P}$. Here, the phases $\alpha$,
$\varphi$ and $\bar{\varphi}$ of $A$, $N$ and
$\bar{N}$ are taken to satisfy the relation
$\alpha+2\varphi+2\bar{\varphi}=\pi$ and $\vert
N\vert$, $\vert\bar{N}\vert$ are assumed equal,
which minimizes the potential. Moreover,
rotating $N$ on the real axis by an appropriate
R transformation, we defined the canonically
normalized real scalar PQ field $\phi=2N$. For
$|A|>4$, this potential has a local minimum at
$\phi=0$ and absolute minima at
\begin{equation}
\langle\phi\rangle^2\equiv f_a^2=\frac{2}{3\lambda}
\left(|A|+\sqrt{|A|^2-12}\right)m_{3/2}m_{\rm P}
\label{eq:fa}
\end{equation}
with $f_a~(>0)$ being the axion decay constant, i.e.
the symmetry breaking scale of ${\rm U}(1)_{\rm PQ}$.
Substituting this vacuum expectation value (VEV)
into the superpotential coupling $\beta N^2h^2/
2m_{\rm P}$ in Eq.~(\ref{W2}), we obtain a $\mu$
term with
\begin{equation}
\mu=\frac{\beta f_a^2}{4m_{\rm P}}\sim m_{3/2},
\label{mu}
\end{equation}
as desired \cite{kn}. Note that the potential
$V_{\rm PQ}$ in Eq.~(\ref{PQ-pot}) should be shifted
\cite{dllr1} by adding to it the constant
\begin{eqnarray}
V_0&=&\frac{1}{108\lambda}\left(|A|+\sqrt{|A|^2-12}
\right)\nonumber \\
& &\times\left[|A|\left(|A|+
\sqrt{|A|^2-12}\right)-24\right]m_{3/2}^3 m_{\rm P},
\label{eq:v0}
\end{eqnarray}
so that it vanishes at its absolute minima.

\section{The PQ field in the early universe}
\label{sec:early}

\par
In the early universe, the PQ potential can acquire
sizable corrections from the SUSY breaking caused
by the presence of a finite energy density
\cite{lyth,crisis,drt95}. In particular, during
inflation and the subsequent inflaton oscillations,
SUSY breaking is transmitted to the PQ system via
its coupling to the inflaton given by the SUGRA
scalar potential. The resulting corrections, whose
scale is set by the Hubble parameter $H$, dominate
over the contributions from hidden sector SUSY
breaking as long as $H\gg m_{3/2}$. For simplicity,
we will ignore the $A$ term type corrections
\cite{drt95}. To leading order, we then just obtain
a correction $\delta m_{\phi}^2$ to the
${\rm mass}^2$ of the curvaton. For a general
K\"{a}hler potential, $\delta m_{\phi}^2\sim H^2$
with either sign possible. However, for specific
(no-scale like) K\"{a}hler potentials, it might be
(partially) cancelled \cite{noscale}. Assuming that
$\delta m_{\phi}^2>0$, we write
\begin{equation}
\delta m_{\phi}^2=\gamma^2 H^2,
\label{eq:effec-mass}
\end{equation}
where $\gamma$ can have different values during
inflation and inflaton oscillations. Actually,
we must assume that, during inflation, $\gamma
\ll 1$ so that the PQ field qualifies as an
effectively massless field which acquires
perturbations from inflation and thus can act
as curvaton. Fortunately, the cancellation of
$\delta m_{\phi}^2$ during inflation can, in
principle, be ``naturally'' arranged to be
exact (see fourth paper in
Ref.~\cite{noscale}). So, for simplicity, we
could take $\gamma=0$ during inflation. On the
other hand, large values of $\gamma$ after the
end of inflation would generically lead
\cite{dllr2} to a drastic reduction of the
density fraction of the PQ field, which thus
again would become unable to play the role of
curvaton. In view of the fact that, in contrast
to the case of inflation, it is not so easy to
achieve exact cancellation of
$\delta m_{\phi}^2$ during inflaton
oscillations, we will only assume that, after
the termination of inflation, $\gamma$ is
somewhat suppressed, say $\sim 0.1$ (or
smaller).

\par
After reheating, the universe is radiation
dominated and, thus, $H\simeq 1/2t\leq 1/
2t_{\rm reh}=\Gamma_{\rm infl}/2$, where $t$ is
the cosmic time and
$t_{\rm reh}=\Gamma_{\rm infl}^{-1}$ the time at
reheating with $\Gamma_{\rm infl}$ being the
inflaton decay width. It is easily seen that, in
this case, $H\ll m_{3/2}$ as a consequence of the
gravitino constraint ($T_{\rm reh}\lesssim 10^9~
{\rm GeV}$) \cite{ekn} on the reheat temperature
$T_{\rm reh}$, which is given by \cite{lectures}
\begin{equation}
T_{\rm reh}=\left(\frac{45}{2\pi^2 g_*}
\right)^{\frac{1}{4}}
(\Gamma_{\rm infl}m_{\rm P})^{\frac{1}{2}},
\label{eq:reheat}
\end{equation}
where $g_*$ is the effective number of degrees
of freedom ($g_*=228.75$ for the MSSM spectrum).
Thus, the SUSY breaking effects from the finite
energy density in the universe are subdominant
compared to the hidden sector SUSY breaking
effects, whose scale is set by $m_{3/2}$.

\par
The PQ potential can acquire temperature
corrections too. During the era of inflaton
oscillations, they originate from the new
radiation \cite{reheat} which emerges from the
decaying inflaton. It has been shown \cite{dllr1},
however, that these corrections are overshadowed
by the SUGRA ones which, in the latest stages of
this era, are, in turn, overshadowed by the terms
from hidden sector SUSY breaking. After reheating,
the temperature corrections are less important
than the ones from the hidden sector as argued in
Refs.~\cite{jean} and \cite{talks}. So, the
temperature corrections to the PQ potential can
be ignored throughout.

\par
We see that, in the early universe, the effective
scalar potential for the PQ field can be taken
to be
\begin{equation}
V_{\rm PQ}^{\rm eff}=V_{\rm PQ}+\frac{1}{2}
\gamma^2 H^2\phi^2+V_0.
\label{full-pot}
\end{equation}
The full effective scalar potential $V$ which is
relevant for our analysis here is obtained by
adding to $V_{\rm PQ}^{\rm eff}$ the potential
for standard SUSY hybrid inflation (see e.g.
Ref.~\cite{lectures}).

\par
The evolution of the field $\phi$ is generally
governed by the classical equation of motion
\begin{equation}
\ddot{\phi}+3H\dot{\phi}+V^{\prime}=0,
\label{field-eqn}
\end{equation}
where overdots and primes denote derivation
with respect to the cosmic time $t$ and the
PQ field $\phi$ respectively. In particular,
Eq.~(\ref{field-eqn}) describes
\cite{communication} the evolution during
inflation of the mean value of $\phi$ in a
comoving region larger than the inflationary
horizon. Actually, this equation starts to be
valid a short time after this region crosses
outside the de Sitter horizon. The mean value
of $\phi$, however, in a region of fixed size
somewhat bigger than the (almost) constant
size of the de Sitter horizon satisfies
\cite{notari} this equation during inflation
provided that it exceeds a certain value
$\phi_Q$ given by
\begin{equation}
V^{\prime}\sim\frac{3H_{\rm infl}^3}{2\pi},
\label{phiQ}
\end{equation}
where $H_{\rm infl}$ is the almost constant
Hubble parameter during inflation.
Therefore, if we require that the value
$\phi_f$ (considered positive without loss of
generality) which is taken at the end of
inflation by the mean $\phi$ in a region of
fixed size somewhat bigger than
$H_{\rm infl}^{-1}$ exceeds $\phi_Q$, we can be
sure \cite{dllr1} that the classical evolution
equation holds for this mean field until the
end of inflation. For values of the mean $\phi$
in a region of fixed size somewhat bigger than
$H_{\rm infl}^{-1}$ which are smaller than
about $\phi_Q$, the random walk executed
\cite{rwalk} by this mean field due to the
quantum perturbations from inflation cannot be
neglected and may overshadow \cite{notari} its
classical motion. Thus, in this case, the
classical equation of motion during inflation
for this mean $\phi$ ceases to be valid. For
reasons to be discussed later, we will not
consider in our analysis values of $\phi_f$
which lie in the quantum regime, i.e. which
are smaller than $\phi_Q$ (see
Secs.~\ref{sec:curvature} and
\ref{sec:evolution}). It should be pointed
out in passing that the requirement of
complete randomization of the mean $\phi$ in
a region of fixed size somewhat bigger than
$H_{\rm infl}^{-1}$ as a consequence of its
quantum perturbations from inflation implies
\cite{prep} an even more stringent bound on
this mean $\phi$ given by
$V\lesssim H^4_{\rm infl}$.

\par
Moreover, as explained in the next section, we
will have to study only values of $\phi_f$ for
which $\phi$ is a slowly rolling field
during the relevant part of inflation (i.e.
during at least the last $50-60$ e-foldings).
It has been shown \cite {dllr1} that, in this
case, the PQ field $\phi$ emerges at the end of
inflation with negligible velocity (i.e.
derivative with respect to cosmic time). Its
subsequent evolution during the matter
dominated era of damped inflaton oscillations
is given by Eqs.~(\ref{full-pot}) and
(\ref{field-eqn}) with $H=2/3t$. One finds
\cite{dllr1} that, depending on the value of
$\phi_f$, the PQ system eventually enters into
a phase of damped oscillations about either the
trivial (local) minimum of $V_{\rm PQ}$ at
$\phi=0$ or one of its PQ (absolute) minima at
$\phi=\pm f_a$. Of course, values of $\phi_f$
leading to the trivial minimum must be excluded.

\par
The damped oscillations of the PQ field continue
even after reheating, where $H$ becomes equal to
$1/2t$. Finally, this field decays via the second
coupling in the superpotential of Eq.~(\ref{W2})
into a pair of Higgsinos provided that their mass
$\mu$ does not exceed half of the mass of the PQ
field (see Ref.~\cite{dllr1})
\begin{equation}
m_{\phi}=\frac{m_{3/2}}{\sqrt{3}}\left(|A|^2-
12\right)^{\frac{1}{4}}\left(|A|+\left(|A|^2-12
\right)^{\frac{1}{2}}\right)^{\frac{1}{2}},
\label{mphi}
\end{equation}
which is independent of the parameter $\lambda$.
The decay time of the PQ field is
$t_{\phi}=\Gamma_{\phi}^{-1}$, where
$\Gamma_{\phi}$ is its decay width, which has
been found \cite{dllr1} to be given by
\begin{equation}
\Gamma_{\phi}=\frac{\beta^2f_a^2}
{8\pi m_{\rm P}^2}m_{\phi}.
\label{Gphi}
\end{equation}
Note that the coherently oscillating PQ field
could evaporate \cite{evap} as a result of
scattering with particles in the thermal bath
before it decays into Higgsinos. However, one
can show \cite{dllr1} that, in the model under
discussion here, this does not happen.

\section{The curvature perturbation}
\label{sec:curvature}

\par
We will consider here only values of $\phi_f$ for
which the PQ field $\phi$ is effectively massless,
i.e. $V^{\prime\prime}\leq H^2$, during (at
least) the last $50-60$ inflationary e-foldings
so that it receives a super-horizon spectrum of
perturbations from inflation
and can act as curvaton. This requirement also
guarantees that $\phi$ is slowly rolling during
the relevant part of inflation. The perturbation
$\delta\phi$ then evolves at subsequent times and,
when $\phi$ settles into damped quadratic
oscillations about the PQ vacua, yields a stable
perturbation in the energy density of this field.
After the PQ field decays, this perturbation is
transferred to radiation, thereby contributing to
the total curvature perturbation. On the other
hand, the radiation, which originates from the
inflaton decay, could carry a curvature
perturbation prior to the curvaton decay too. It
actually inherits the curvature perturbation of
the inflaton. Contrary to the standard curvaton
hypothesis \cite{curv1}, we make here the more
natural assumption that this perturbation is
non-zero and, thus, also contributes to the total
curvature perturbation.

\par
The scalar part of the metric perturbation for
a flat universe can be written (using the
notation of Ref.~\cite{thesis}), in all
generality, as follows (for reviews of the
gauge invariant theory of cosmological
perturbations, see e.g. Ref.~\cite{gipt}):
\begin{eqnarray}
\delta g_{\mu \nu}d x^\mu d x^\nu &=&
-2Adt^2+2aB_{,i}dtdx^i\nonumber \\
& &+a^2\left(2C\delta_{ij}
+E_{,ij}\right)dx^idx^j,
\label{metric}
\end{eqnarray}
where $\mu,\nu=0,1,2,3$ and $i,j=1,2,3$. Here,
$x^0=t$ is the physical time, $x^i$ ($i=1,2,3$)
are the comoving spatial coordinates,
$Y_{,i}\equiv\partial Y/\partial x^i$
($i=1,2,3$) and $\delta_{ij}$ denotes the
Kronecker delta. The dimensionless parameter
$a$ is the scale factor of the universe, which
is normalized to unity at the present cosmic
time. We define the following gauge invariant
quantities:
\begin{eqnarray}
\zeta&\equiv &C-H\frac{\delta\rho}{\dot{\rho}},
\label{zetagen}
\\
\mathcal{R}_{\rm rad}&\equiv &-C-H(B-a^2v),
\\
\Phi &\equiv &-C-H\left(B-a^2\dot{E}\right).
\end{eqnarray}
Here, $\rho=-T_0^{~0}$ is the total energy
density in the universe with $T_\mu^{~\nu}$
being the energy momentum tensor, $\delta\rho
=-\delta T_0^{~0}$ is the total density
perturbation, and $v$ describes the spatial
perturbation in the 4-velocity $u^\mu$ of an
observer comoving with the total fluid, i.e.
$v^{,i}\equiv -\delta u^i/u^0$. The variable
$\zeta$ represents the curvature perturbation
on hypersurfaces of uniform density,
$\mathcal{R}_{\rm rad}$ is the curvature
perturbation in the (total matter) comoving
gauge (up to the sign), while $\Phi$ is the
Bardeen potential, which is the curvature
perturbation (up to the sign) in the
longitudinal gauge. These three quantities are
related by
\begin{equation}
\zeta=-\mathcal{R}_{\rm rad}-\frac{2\rho}
{9(\rho+p)}\left(\frac{k}{Ha}\right)^2\Phi,
\end{equation}
where we have used the time-time component of
the Einstein equation and $p$ is the total
pressure of the universe (i.e. $T_{ij}= p
\delta_{ij}$). On super-horizon scales, $k\ll
Ha$, the second term on the right hand side of
this equation is negligible, and we thus have
$\zeta=-\mathcal{R}_{\rm rad}$.

\par
Using Eq.~(\ref{zetagen}), the total curvature
perturbation \cite{bst} in the flat slicing
gauge (defined by setting $C=E=0$ in
Eq.~(\ref{metric})) is given by
\begin{equation}
\zeta=\frac{\delta\rho}{3(\rho+p)}.
\end{equation}
After the curvaton decay, it becomes
\cite{curv3}
\begin{equation}
\zeta=(1-f)\zeta_i+f\zeta_c,
\label{zeta}
\end{equation}
where $\zeta_i=\delta\rho_r/4\rho_r$ and
$\zeta_c=\delta\rho_\phi/3\rho_\phi$ are the
partial curvature perturbations on spatial
hypersurfaces of constant curvature from the
inflaton and the curvaton respectively at the
curvaton decay with $\rho_r$ and $\rho_\phi$
being the radiation and $\phi$ energy densities
respectively, and $\delta\rho_r$ and $\delta
\rho_\phi$ the corresponding perturbations.
Also,
\begin{equation}
f=\frac{3\rho_{\phi}}{3\rho_{\phi}+4\rho_r}
\label{eq:f}
\end{equation}
evaluated at the time of the curvaton decay. Here,
we assume that
the amplitude of the oscillating $\phi$ has been
sufficiently reduced so that the potential can be
approximately considered as quadratic. Actually,
as shown in Ref.~\cite{dllr2}, this must
necessarily happen before the curvaton decays.
The oscillating curvaton field then behaves like
pressureless matter and $\zeta_c=2\delta\phi_0/
3\phi_0$, where $\phi_0$ is the amplitude of the
oscillations and $\delta\phi_0$ the perturbation
in this amplitude originating from the
perturbation $\delta\phi_f$ in the value $\phi_f$
of $\phi$ at the end of inflation.

\par
The comoving curvature perturbation
$\mathcal{R}_{\rm rad}$, for super-horizon
scales, is given by
\begin{equation}
\mathcal{R}_{\rm rad}=(1-f)A_i\left(\frac{k}{H_0}
\right)^{\nu_i}\hat{a}_i+
fA_c\left(\frac{k}{H_0}\right)^{\nu_c}\hat{a}_c,
\label{R}
\end{equation}
where $k$ is the comoving (present physical)
wave number, $H_0$
is the present value of the Hubble parameter
and $\hat{a}_i$, $\hat{a}_c$ are independent
normalized Gaussian random variables. Also,
$A_i$ and $A_c$ are, respectively, the
amplitudes of $-\zeta_i$ and $-\zeta_c$ at the
present horizon scale (i.e. at $k=H_0$), and
$\nu_i$ and $\nu_c$ are the spectral tilts of
the inflaton and curvaton respectively, which
are related to the corresponding spectral
indices $n_i$ and $n_c$ by $n_i=1+2\nu_i$ and
$n_c=1+2\nu_c$. We do not consider running of
the spectral indices, since this is negligible
in our model.

\par
The amplitude $A_i$ of the partial curvature
perturbation from the inflaton is found \cite{hier}
to be
\begin{equation}
A_i=\left(\frac{2N_{Q}}{3}\right)^{\frac{1}{2}}
\left(\frac{M}{m_{\rm P}}\right)^2x_Q^{-1}y_Q^{-1}
\Lambda(x_Q^2)^{-1}
\label{ai}
\end{equation}
with
\begin{equation}
\Lambda(z)=(z+1)\ln(1+z^{-1})+(z-1)\ln(1-z^{-1})~,
\label{lambda}
\end{equation}
\begin{equation}
y_Q^2=\int_{x_f^2}^{x_Q^2}\frac{dz}{z}
\Lambda(z)^{-1},~y_Q\geq 0~.
\label{yq}
\end{equation}
Here, $N_{Q}$ is the number of e-foldings suffered
by our present horizon scale during inflation,
$x_Q=S_Q/M$ with $S_Q$ being the value
of $\vert S\vert$ when our present horizon scale
crosses outside the inflationary horizon, and
$x_f=S_f/M$ with $S_f$ being the value
of $\vert S\vert$ at the end of inflation.

\par
In our model, the slow-roll parameters for the
inflaton as functions of $\vert S\vert$ are given
by \cite{lectures}
\begin{equation}
\epsilon_i=\left(\frac{\kappa^2m_{\rm P}}{8\pi^2M}
\right)^2z\Lambda(z)^2,
\label{epsiloni}
\end{equation}
\begin{eqnarray}
\eta_i&=&2\left(\frac{\kappa m_{\rm P}}{4\pi M}
\right)^2\Big[(3z+1)\ln(1+z^{-1})
\nonumber\\
& &+(3z-1)\ln(1-z^{-1})\Big],
\label{etai}
\end{eqnarray}
where $z=x^2$ with $x=\vert S\vert/M$. In the
presence of the curvaton, however, one can show
that the slow-roll conditions (for $\gamma
\rightarrow 0$) take the form $\epsilon,
\vert\eta_i\vert\leq 1$, where
\begin{equation}
\epsilon\equiv -\frac{\dot H}{H^2}=\epsilon_i+
\epsilon_c
\label{epsilon}
\end{equation}
with
\begin{equation}
\epsilon_c=\frac{1}{2}
m_{\rm P}^2\left(\frac{V^{\prime}}{V}\right)^2,
\label{epsilonc}
\end{equation}
and $x_f$ is given by the value of $x$ for
which these conditions are saturated. Actually,
as it turns out, $x_f$ corresponds to
$\eta_i=-1$. For $\kappa\ll 1$, the slow-roll
conditions are violated only extremely close to
the critical point at $x=1$
($\vert S\vert=S_c$). So, inflation continues
practically until this point is reaches and,
following Ref.~\cite{hier}, we can put $x_f=1$
in Eq.~(\ref{yq}). However, for larger values
of the parameter $\kappa$, inflation can
terminate well before reaching the instability
point.

\par
Finally, $\kappa$, $N_Q$ are given by
\cite{lectures}
\begin{equation}
\kappa=\frac{2\pi}{\sqrt{N_Q}}
~y_Q~\frac{M}{m_{\rm P}},
\label{kappa}
\end{equation}
\begin{equation}
N_Q\simeq 55.9+\frac{2}{3}\ln\frac{
\kappa^{\frac{1}{2}}M}{10^{15}~{\rm GeV}}
+\frac{1}{3}\ln\frac{T_{\rm reh}}
{10^9~{\rm GeV}}
\label{NQ}
\end{equation}
(for $H_0\simeq 72~\rm{km}~\rm{sec}^{-1}~
\rm{Mpc}^{-1}$). The spectral index for the
inflaton is $n_i=1-6\epsilon+2\eta_i$ (for
$\gamma\rightarrow 0$), where $\epsilon$ and
$\eta_i$ are evaluated at the time when
our present horizon scale crosses outside
the inflationary horizon. Note that $\epsilon$
enters in this formula, not $\epsilon_i$ as in
the case of pure inflaton. However,
$\epsilon_c$ is normally much smaller than
$\epsilon_i$ which, in turn, is negligible
compared to $\vert\eta_i\vert$. As a
consequence, $n_i\simeq 1+2\eta_i$.

\par
We now calculate the amplitude $A_c$ of the
partial curvature perturbation from the
curvaton $\phi$. This originates from the
perturbation $\delta\phi_*=(H_*/2\pi)\hat{a}_c$
acquired by $\phi$ from inflation when our
present horizon scale crosses outside the de
Sitter horizon
($H_*$ is the inflationary Hubble parameter at
that moment). In order to find the evolution of
this perturbation during the subsequent part of
inflation and estimate its value $\delta\phi_f$
at the end of inflation, we must consider the
equation of motion for $\phi$ during inflation
(see Eq.~(\ref{field-eqn})). In the slow-roll
approximation, which is assumed to hold for the
curvaton too, this equation reads
\begin{equation}
3H\dot\phi+V^{\prime}=0.
\label{slow}
\end{equation}
Taking a small perturbation
$\delta\phi$ of $\phi$, Eq.~(\ref{slow}) gives
\begin{equation}
3H\delta\dot\phi+3H^{\prime}\delta\phi\dot\phi+
V^{\prime\prime}\delta\phi=0.
\label{dphi1}
\end{equation}
Substituting $\dot\phi$ from Eq.~(\ref{slow}) and
using the Friedmann equation $3H^2m_{\rm P}^2=V$,
Eq.~(\ref{dphi1}) becomes
\begin{equation}
\delta\dot\phi+H(-\epsilon_c+\eta_c)\delta\phi=0,
\label{dphi2}
\end{equation}
where
\begin{equation}
\eta_c=m_{\rm P}^2\frac{V^{\prime\prime}}{V}.
\label{etac}
\end{equation}
Integration of Eq.~(\ref{dphi2}) from the cosmic
time $t_*$ when our present horizon scale crossed
outside the inflationary horizon until the end of
inflation (at time $t_f$) yields
\begin{equation}
\delta\phi_f=\frac{H_*}{2\pi}\,\hat{a}_c
\exp{\int_0^{N_Q}(\epsilon_c-\eta_c)dN},
\label{dphif}
\end{equation}
where we used the relation $dN=-Hdt$ for the
number of e-foldings $N(k)=N_Q+\ln(H_0/k)$
suffered by the scale which corresponds to the
comoving wave number $k$ during hybrid inflation.

\par
For each value $\phi_f~(>0)$ of the curvaton
field at the end of inflation, we construct the
perturbed field $\phi_f+\delta\phi_f$. We then
follow the evolution of $\phi_f$ and $\phi_f+
\delta\phi_f$ until the time $t_\phi$ of the
curvaton decay and evaluate the value of
$\delta\rho_\phi/\rho_\phi$ at this time. The
amplitude $A_c$ of the partial curvature
perturbation from the curvaton is given by
\begin{equation}
A_c\,\hat{a}_c=\frac{1}{3}\frac{\delta\rho_\phi}
{\rho_\phi}.
\label{Ac}
\end{equation}
We have found numerically that the perturbation
$\delta\phi_0$ in the amplitude of the oscillating
curvaton at $t_\phi$ is proportional to
$\delta\phi_f$. So $\zeta_c$ has the same spectral
tilt as $\delta\phi_f$, which can be found from
Eq.~(\ref{dphif}):
\begin{equation}
\nu_c\equiv\frac{d\ln A_c}{d\ln k}=-\epsilon-
\epsilon_c+\eta_c,
\label{nuc}
\end{equation}
where we used the relation $d\ln k=Hdt$, and
$\epsilon$, $\epsilon_c$ and $\eta_c$ are
evaluated at $t_*$. (Note that a similar
formula has been derived in the first paper
in Ref.~\cite{curv2}, but without the
$-\epsilon_c$ term in the right hand side.)
The spectral index for the
curvaton is then $n_c=1-2\epsilon-2\epsilon_c+
2\eta_c$ which, in most cases, reduces to $n_c
\simeq 1+2\eta_c$ since we typically have
$\epsilon_i,\epsilon_c\ll\vert\eta_c\vert$.

\par
It should be pointed out that, in deriving
Eq.~(\ref{nuc}), we assume that, during
inflation, the mean value of $\phi$ in any
region of fixed size (somewhat bigger than)
$H_{\rm infl}^{-1}$ is initially the same and
follows the classical equation of motion
(Eq.~(\ref{field-eqn})). This mean field sets
the (initial) value of the mean $\phi$ in any
comoving region at the time this region exits
the de Sitter horizon. Thus, if it satisfies
the above requirements, we can be sure that,
at any given time during inflation, the
resulting mean $\phi$ in each comoving region
that already exited the horizon is practically
independent from the size of this region. So
the mean $\phi$ in a fixed region of size
$H_{\rm infl}^{-1}$ or in any comoving volume
(after horizon exit) is described by the same
(single-valued) function of the cosmic time
and evolves classically (in fact, it rolls
down slowly in our case). The derivation with
respect to time of the logarithm of the
amplitude of $\delta\phi_f$, which is given by
Eq.~(\ref{dphif}), is then straightforward
leading to Eq.~(\ref{nuc}). On the contrary,
if the mean field in a fixed region of size
somewhat bigger than $H_{\rm infl}^{-1}$
executes a random walk with step of amplitude
$H_{\rm infl}/2\pi$ per Hubble time, the
calculation of the spectral tilt becomes less
clear. So, we decided to exclude the quantum
regime (i.e. the region $\phi_f<\phi_Q$) from
our analysis (see Sec.~\ref{sec:evolution}). It
is true, however, that it is by no means
necessary to avoid the random walk behavior at
all times during inflation. Indeed, only
cosmological scales corresponding to a few
e-foldings after the exit of our present
horizon scale from the de Sitter horizon are
relevant. For simplicity though, we exclude all
the quantum regime so that no random motion of
the mean $\phi$ in a region of size
$H_{\rm infl}^{-1}$ is encountered during
inflation. This, as we will see, has no
influence on our results.

\section{The isocurvature perturbation}
\label{sec:isocurvature}

\par
After the termination of inflation, the
inflaton performs damped oscillations about
the SUSY vacuum and eventually decays into
light particles reheating the universe. At
reheating, gravitinos are
thermally produced besides other particles.
They only decay, though, well after the big
bang nucleosynthesis (BBN) since they have
very weak couplings. Each decaying gravitino
yields one sparticle subsequently turning
into the LSP, which is stable. These LSPs
survive until the present time contributing
to the relic abundance of CDM in the universe.
For simplicity, we assume that the thermally
produced LSPs can be neglected, which holds
in many cases. So, all the relic LSPs come
solely from the decaying gravitinos. Baryons
can be produced via a primordial leptogenesis
\cite{lepto} which can occur \cite{leptoinf}
at reheating.

\par
We see that both the LSPs and the baryons
originate from reheating. Their partial
curvature perturbations, $\zeta_{\rm LSP}$
and $\zeta_B$ respectively, should thus
coincide with the partial curvature
perturbation of the radiation which emerges
from the inflaton decay, i.e.
\begin{equation}
\zeta_{\rm LSP}=\zeta_B=\zeta_i.
\label{zetaLSP}
\end{equation}
The isocurvature perturbation of the LSPs and
the baryons is then given by
\begin{equation}
\mathcal{S}_{{\rm LSP}+B}\equiv 3(
\zeta_{{\rm LSP}+B}-\zeta)=3f(\zeta_i-\zeta_c),
\label{SLSP1}
\end{equation}
where $\zeta_{{\rm LSP}+B}=\zeta_{\rm LSP}=
\zeta_B$ is the partial curvature perturbation
of the LSPs and the baryons and we used
Eq.~(\ref{zeta}). Here, we assume
that the curvature perturbation in radiation
($\zeta_\gamma$) practically coincides with the
total curvature perturbation. This corresponds to
a negligible neutrino isocurvature perturbation,
which is \cite{curv3} the case provided that, as
in our model, leptogenesis takes place well
before the curvaton decays or dominates the
energy density. Applying the definitions which
follow Eq.~(\ref{R}), Eq.~(\ref{SLSP1}) takes the
form
\begin{equation}
\mathcal{S}_{{\rm LSP}+B}=-3fA_i\left(
\frac{k}{H_0}\right)^{\nu_i}\hat{a}_i+3fA_c\left(
\frac{k}{H_0}\right)^{\nu_c}\hat{a}_c.
\label{SLSP2}
\end{equation}

\par
Our model contains axions which can also
contribute to the CDM of the universe. They are
produced at the QCD phase transition, which
occurs well after the curvaton decay. They carry
an isocurvature perturbation, which is completely
uncorrelated with the curvature perturbation and
can be written as
\begin{equation}
\mathcal{S}_a=A_a\left(\frac{k}{H_0}
\right)^{\nu_a}\hat{a}_a,
\label{Sa}
\end{equation}
where $A_a$ is its amplitude at the present
horizon scale, $\nu_a$ is the corresponding
spectral tilt (yielding the spectral index
$n_a=1+2\nu_a$) and $\hat{a}_a$ is a normalized
Gaussian random variable which is independent
from $\hat{a}_i$ and $\hat{a}_c$.

\par
The amplitude $A_a$ is given by \cite{dllr1}
\begin{equation}
A_a=\frac{H_*}{\pi\vert\theta\vert\phi_*},
\label{Aa}
\end{equation}
where $\theta$ is the so-called initial
misalignment angle, i.e. the phase of the complex
PQ field during (and at the end of) inflation and
$\phi_*$ is the value of $\phi$ at $t_*$. In our
case, the angle $\theta$ lies \cite{dllr1} in the
interval $[-\pi/6,\pi/6]$ with all values in it
being equally probable. It is determined by
considering the total CDM abundance
$\Omega_{\rm CDM}h^2$ which, in our model, is
the sum of the relic abundance \cite{km}
\begin{equation}
\Omega_{\rm LSP}h^2\simeq 0.0074\left(\frac
{m_{\rm LSP}}{200~{\rm GeV}}\right)\left(
\frac{T_{\rm reh}}{10^{9}~{\rm GeV}}\right)
\label{OmegaLSP}
\end{equation}
of the LSPs coming from the gravitino decays
and the relic axion abundance \cite{axion}
\begin{equation}
\Omega_ah^2\simeq\theta^2\left(\frac{f_a}
{10^{12}~{\rm GeV}}\right)^{1.175}.
\label{Omegaa}
\end{equation}
Here $\Omega_j=\rho_j/\rho_c$ with $\rho_j$
being the present energy density of the $j$th
species and $\rho_c$ the present critical
energy density of the universe. Furthermore,
$m_{\rm LSP}$ is the LSP mass and
$H_0=100h~\rm{km}~\rm{sec}^{-1}~\rm{Mpc}^{-1}$
the present Hubble parameter. In the numerical
calculation of $M$ and the amplitudes $A_c$
and $A_a$ for given $A_i$ and $\phi_f$ (see
Secs.~\ref{sec:evolution} and \ref{sec:Cell}),
we will fix $h\simeq 0.72$, which is its
best-fit value from the Hubble space telescope
(HST) \cite{hst}. (In the full Monte Carlo (MC)
CMBR analysis, however, we do allow $h$ to vary
as detailed in Sec.~\ref{sec:setup}.) This
approximation has very little influence on the
accuracy of the results (see the remarks at the
end of Sec.~\ref{sec:Cell}). Note that, in
deriving the estimate for $\Omega_a h^2$ in
Eq.~(\ref{Omegaa}), we applied the simplifying
assumptions of Ref.~\cite{dllr1}.

\par
The spectral tilt $\nu_a$ is evaluated by
observing that the potential of the axion field
remains flat until the QCD transition is reached.
So, there is no evolution of $\theta$ and its
perturbation $\delta\theta=(H_*/2\pi\phi_*)
\hat{a}_a$ after crossing outside the
inflationary horizon and
until the onset of axion oscillations. The
axion isocurvature perturbation, which is
\cite{dllr1} equal to $2\delta\theta/\theta$,
depends on the scale only through $H_*/\phi_*$.
We find
\begin{equation}
\nu_a=-\epsilon+\frac{m_{\rm P}}{\phi_*}
\frac{V^\prime}{V}m_{\rm P}=
-\epsilon+\frac{m_{\rm P}}{\phi_*}
\sqrt{2\epsilon_c},
\label{nua}
\end{equation}
where Eqs.~(\ref{epsilonc}) and (\ref{slow})
were used, and the slow-roll parameters
$\epsilon$ and $\epsilon_c$ are evaluated at
$t_*$. In view of the fact that $\epsilon$ is
normally negligible, Eq.~(\ref{nua}) reduces
to $\nu_a\simeq m_{\rm P}\sqrt{2\epsilon_c}/
\phi_*$ and $n_a\simeq 1+2m_{\rm P}
\sqrt{2\epsilon_c}/\phi_*$.

\par
Combining Eqs.~(\ref{SLSP2}) and (\ref{Sa}), we
find that the total isocurvature perturbation is
\cite{dllr1}
\begin{equation}
\mathcal{S}_{\rm rad}=\frac{\Omega_{{\rm LSP}+B}}
{\Omega_m}\mathcal{S}_{{\rm LSP}+B}+\frac{\Omega_a}
{\Omega_m}\mathcal{S}_a,
\label{total}
\end{equation}
where we used the definitions $\Omega_{{\rm LSP}
+B}\equiv\Omega_{\rm LSP}+\Omega_B$ and $\Omega_m
\equiv\rho_m/\rho_c=\Omega_{{\rm LSP}+B}+\Omega_a$
with $\rho_m$ being the total matter density at
present.

\section{The CMBR power spectrum}
\label{sec:spectrum}

\par
In order to calculate the expected total CMBR
angular power spectrum $C_\ell$ in our model, we
must first specify the various contributions to
the amplitude of the total adiabatic and
isocurvature perturbation as well as the
cross correlation between these two perturbations.
In the following, all the amplitudes are referred
to the pivot scale $k_{\rm P}$, for which we use
the customary value $k_{\rm P}=0.05~{\rm Mpc^{-1}}$
\cite{pivot}. We thus define the amplitudes
of the partial curvature perturbations from the
inflaton ($A_{{\rm P},i}$) and the curvaton
($A_{{\rm P},c}$), and the amplitude of the
isocurvature perturbation in the axions
($A_{{\rm P},a}$) at $k=k_{\rm P}$:
\begin{eqnarray*}
A_{{\rm P},i}=A_i\left(\frac{k_{\rm P}}{H_0}
\right)^{\nu_i},~
A_{{\rm P},c}=A_c\left(\frac{k_{\rm P}}{H_0}
\right)^{\nu_c},
\end{eqnarray*}
\begin{equation}
A_{{\rm P},a}=A_a\left(\frac{k_{\rm P}}{H_0}
\right)^{\nu_a},
\label{AP}
\end{equation}
where the amplitudes $A_i$, $A_c$ and $A_a$ at
$k=H_0$ are given in Eqs.~(\ref{ai}),  (\ref{Ac})
and (\ref{Aa}) respectively.

\par
From Eq.~(\ref{R}), we find that the amplitude
squared of the adiabatic perturbation at the
pivot scale is given by
\begin{equation}
R^2=\langle\mathcal{R}_{\rm rad}
\mathcal{R}_{\rm rad}\rangle=R_i^2+R_c^2,
\label{R2}
\end{equation}
where $\mathcal{R}_{\rm rad}$ is evaluated at
$k=k_{\rm P}$, and the inflaton ($R_i^2$) and
curvaton ($R_c^2$) contributions to this amplitude
squared are
\begin{equation}
R_i^2=(1-f)^2A_{{\rm P},i}^2~~{\rm and}~~R_c^2=
f^2A_{{\rm P},c}^2.
\label{Ric}
\end{equation}
The curvaton fractional contribution
to the amplitude of the adiabatic perturbation is
defined as follows:
\begin{equation}
F^{\rm ad}_c=\frac{R_c}{R}.
\label{Fc}
\end{equation}

\par
The amplitude squared of the isocurvature
perturbation at $k_{\rm P}$ is found from
Eq.~(\ref{total}) to be
\begin{equation}
S^2=\langle\mathcal{S}_{\rm rad}
\mathcal{S}_{\rm rad}\rangle=S_i^2+S_c^2+S_a^2,
\label{S2}
\end{equation}
where
\begin{eqnarray*}
S_i^2=9f^2\left(\frac{\Omega_{{\rm LSP}+B}}
{\Omega_m}\right)^2A_{{\rm P},i}^2,
\end{eqnarray*}
\begin{equation}
S_c^2=9f^2\left(\frac{\Omega_{{\rm LSP}+B}}
{\Omega_m}\right)^2A_{{\rm P},c}^2~~{\rm and}~~
S_a^2=\left(\frac{\Omega_a}{\Omega_m}\right)^2
A_{{\rm P},a}^2
\label{Sica}
\end{equation}
are, respectively, the inflaton, curvaton and
axion contributions to this amplitude squared.

\par
The cross correlation between the adiabatic and
isocurvature perturbation at the pivot scale
$k_{\rm P}$ is
\begin{equation}
C=\langle\mathcal{R}_{\rm rad}
\mathcal{S}_{\rm rad}\rangle=C_i+C_c,
\label{C}
\end{equation}
where
\begin{equation}
C_i=-R_iS_i~~{\rm and}~~C_c=R_cS_c
\label{Cic}
\end{equation}
are the contributions from the inflaton and the
curvaton respectively. Observe that the axions
do not contribute to the cross correlation (and
the amplitude of the adiabatic perturbation).
Also, note that the isocurvature perturbation
from the inflaton is fully anti-correlated with
the corresponding adiabatic perturbation,
whereas the isocurvature perturbation from the
curvaton is fully correlated with the adiabatic
perturbation from it. So the overall correlation
is mixed. It is thus useful to define \cite{iso3}
the dimensionless cross correlation $\cos\Delta$
as a measure of the correlation between adiabatic
and isocurvature perturbations and the
entropy-to-adiabatic ratio $B$ at $k_{\rm P}$:
\begin{equation}
\cos\Delta=\frac{C}{RS}~~{\rm and}~~B=\frac{S}{R}.
\label{cosDB}
\end{equation}

\par
The total CMBR angular power spectrum is given,
in the notation of Ref.~\cite{thesis}, by the
superposition
\begin{equation}
C_\ell=C_\ell^{\rm ad}+C_\ell^{\rm is}+
C_\ell^{\rm cc},
\label{Cell}
\end{equation}
where
\begin{equation}
C_\ell^{\rm ad}=R_i^2C_\ell^{{\rm ad},n_i}+
R_c^2C_\ell^{{\rm ad},n_c},
\label{Cellad}
\end{equation}
\begin{equation}
C_\ell^{\rm is}=S_i^2C_\ell^{{\rm is},n_i}+
S_c^2C_\ell^{{\rm is},n_c}+
S_a^2C_\ell^{{\rm is},n_a},
\label{Cellis}
\end{equation}
\begin{equation}
C_\ell^{\rm cc}=C_iC_\ell^{{\rm cc},n_i}+
C_cC_\ell^{{\rm cc},n_c}.
\label{Cellcc}
\end{equation}
The above relations hold for the TT,
E-polarization (EE) and TE cross correlation
angular power spectra.

\par
The TT power spectrum on large angular scales
(i.e. for $\ell\lesssim 20$), can be approximated
by (see e.g. Ref.~\cite{thesis})
\begin{eqnarray}
C^{\rm TT}_\ell&=&\frac{2\pi^2}{25}
\big[(R_i^2+4S_i^2-4C_i)f(n_i,\ell)
\nonumber\\
& &+(R_c^2+4S_c^2-4C_c)f(n_c,\ell)
\nonumber\\
& &+4S_a^2f(n_a,\ell)\big],
\label{Cobe}
\end{eqnarray}
where
\begin{equation}
f(n,\ell)=(\eta_0k_{\rm P})^{1-n}
\frac{\Gamma(3-n)
\Gamma(\ell-\frac{1}{2}+\frac{n}{2})}{2^{3-n}
\Gamma^2(2-\frac{n}{2})\Gamma(\ell+\frac{5}{2}
-\frac{n}{2})}
\label{fnell}
\end{equation}
with $\eta_0=2H_0^{-1}\simeq 8.33\times 10^3~
{\rm Mpc}$ being the value of conformal time in
the present (matter dominated) universe. The
derivation of Eq.~(\ref{Cobe}) assumes that the
universe is completely matter dominated at the
moment of recombination, and further neglects
the late integrated Sachs-Wolfe (ISW) effect
from the cosmological constant. Therefore this
expression is accurate to about $10-20\%$ and
gives an order of magnitude estimate only.
For scale-invariant spectra, $f(n=1,\ell)=1/
\pi\ell(\ell+1)$ and the Sachs-Wolfe (SW)
plateau is flat. Notice from Eq.~(\ref{Cobe})
that a positively correlated perturbation (as
the perturbation from the curvaton) displays
less power on large angular scales, because
the cross correlation term subtracts power and
can partially cancel the isocurvature
contribution. On the contrary, a negatively
correlated perturbation (such as the one from
the inflaton) presents larger power on the
COBE scales and, therefore, can be more easily
constrained. The COBE measurements give
\cite{cobe10} the following central value
for $\ell=10$:
\begin{equation}
\ell(\ell+1)C^{\rm TT}_\ell/2\pi\big
\vert_{\ell=10}\approx 1.05\times 10^{-10}.
\label{ell10}
\end{equation}
We will apply this COBE constraint with
$C^{\rm TT}_\ell$ evaluated from
Eq.~(\ref{Cobe}) only as an indicative
(approximate) criterion to get a first rough
feeling on the possible compatibility of our
model with the data.

\section{Numerical calculations and results}
\label{sec:numerics}

\subsection{The cosmological evolution of
$\phi$}
\label{sec:evolution}

\par
\begin{figure}[tb]
\centering
\includegraphics[width=0.83\linewidth]{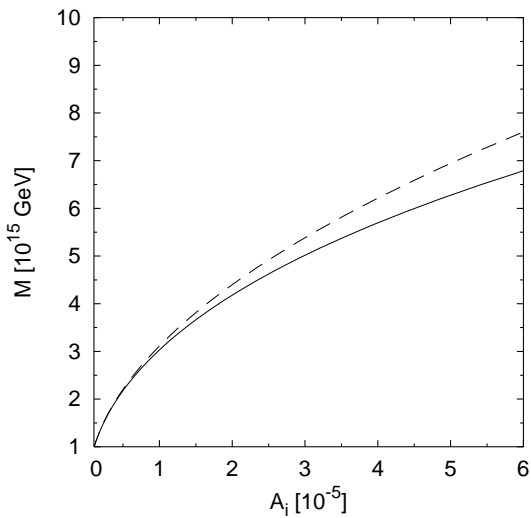}
\caption{The mass parameter $M$ versus $A_i$
for $\kappa=3\times 10^{-3}$ (solid line) or
$3\times 10^{-2}$ (dashed line).}
\label{fig:M}
\end{figure}
\begin{figure}[tb]
\centering
\includegraphics[width=0.83\linewidth]
{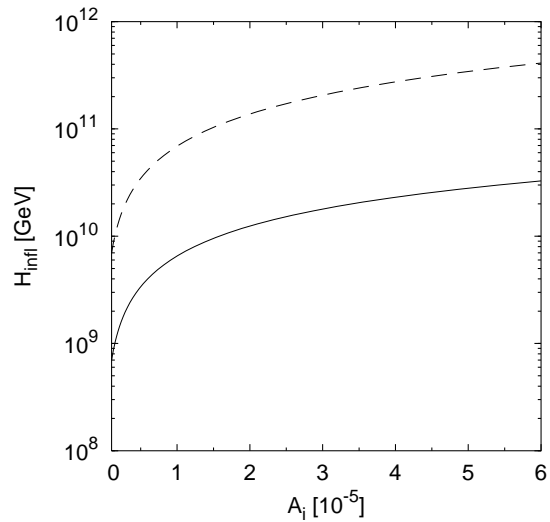}
\caption{The inflationary Hubble parameter
$H_{\rm infl}$ versus $A_i$ for
$\kappa=3\times 10^{-3}$ (solid line) or
$3\times 10^{-2}$ (dashed line).}
\label{fig:Hinf}
\end{figure}

We are now ready to proceed to the numerical
study of the evolution of the PQ field during
and after inflation. To this end, we fix the
parameter $\kappa$ in the superpotential
$\delta W_1$ in Eq.~(\ref{W1}) for standard
hybrid inflation. Then, for any given value of
the amplitude $A_i$ of the partial curvature
perturbation from the inflaton, we solve
Eqs.~(\ref{ai}), (\ref{kappa}) and (\ref{NQ}),
where $x_f$ is the solution of $\eta_i=-1$
with $\eta_i$ given by Eq.~(\ref{etai}) and
$T_{\rm reh}$, which enters Eq.~(\ref{NQ}), is
taken equal to $10^9~{\rm GeV}$ by saturating
the gravitino bound \cite{ekn}. We thus
determine the mass parameter $M$, which is the
magnitude of the VEV breaking the $G_{LR}$
symmetry. Subsequently, we find the (almost
constant) inflationary Hubble parameter
$H_{\rm infl}=\kappa M^2/\sqrt{3}m_{\rm P}$.
The parameters $M$ and $H_{\rm infl}$ as
functions of the amplitude $A_i$ are shown in
Figs.~\ref{fig:M} and \ref{fig:Hinf}
respectively for $\kappa=3\times 10^{-3}$
(solid line) or $3\times 10^{-2}$ (dashed
line). Note that much smaller values of
$\kappa$ would be considered unnatural. On the
other hand, much bigger $\kappa$'s would push
inflation to higher values of the inflaton
field, where SUGRA corrections become important
and may ruin \cite{lyth} inflation.

\par
For any given $A_i$, we chose a value $\phi_f$
of $\phi$ at the end of inflation, which takes
place at cosmic time $t_f=2/3H_{\rm infl}$. We
then solve the classical equation of motion
for $\phi$ during inflation in the slow-roll
approximation going backwards in time
($t\leq t_f$) by taking $m_{3/2}=300~
{\rm GeV}$ (which is \cite{wang} approximately
its minimal value in the
constrained MSSM with Yukawa unification
\cite{ananth}), $\vert A\vert=5$ and a fixed
value for $\lambda$ ($\sim 10^{-4}$) in the PQ
potential. Note that much bigger values of
$\vert A\vert$ (recall that $\vert A\vert>4)$
or much smaller values of $\lambda$ would not
only be unnatural, but also would lead to an
enhancement of the axion decay constant $f_a$
(see Eq.~(\ref{eq:fa})) and thus require
unnaturally small misalignment angles $\theta$
to achieve the observed total CDM abundance
(see Eq.~(\ref{Omegaa})). On the other hand,
much bigger values of $\lambda$ would reduce
$f_a$ leading to an unacceptably small CDM
abundance. Finally, the parameter $\gamma$
during inflation is put equal to zero for
simplicity (see Sec.~\ref{sec:early}). We find
that, as we move backwards in time, the value
of $\phi$ increases and becomes infinite at a
certain moment. Also, the number of e-foldings
elapsed from this moment of time until the end
of inflation is finite providing an upper bound
$N_{\rm max}$ on the number of e-foldings which
is compatible with our boundary condition
$\phi=\phi_f$ at $t_f$.

\par
In order to get a (rough) understanding of this
behavior, we approximate the derivative of the
potential $V$ with respect to $\phi$ by
$V^{\prime}\simeq 3\lambda^2\phi^5/16
m_{\rm P}^2$, which holds for $\phi\rightarrow
\infty$. The slow-roll equation for $\phi$
during inflation can then be solved analytically
and yields
\begin{equation}
\phi\simeq\frac{\phi_f}
{\left(1+\frac{\lambda^2\phi_f^4}
{4m_{\rm P}^2H_{\rm infl}}\Delta t
\right)^{\frac{1}{4}}},
\label{analytic}
\end{equation}
where $\Delta t=t-t_f\leq 0$. It is obvious
from this equation that, as $\Delta t
\rightarrow\Delta t_{\rm min}\equiv
-4m_{\rm P}^2H_{\rm infl}/\lambda^2\phi_f^4$,
$\phi\rightarrow\infty$. This implies that
the maximal number of e-foldings which is
allowed for a given value of $\phi_f$ is
$N_{\rm max}\simeq -H_{\rm infl}\Delta
t_{\rm min}=4m_{\rm P}^2H_{\rm infl}^2/
\lambda^2\phi_f^4$. Needless to say that no
approximation for $V^{\prime}$ is used when
we actually calculate $N_{\rm max}$.

\par
\begin{figure}[tb]
\centering
\includegraphics[width=0.83\linewidth]
{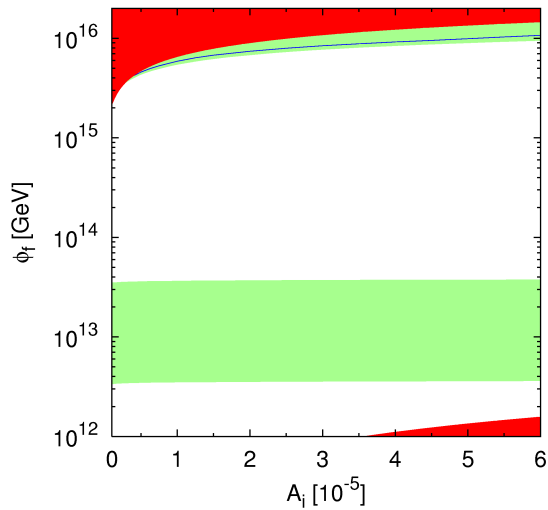}
\caption{The two green/lightly shaded bands
in the $A_i-\phi_f$ plane which lead to the
PQ vacua at $t_{\phi}$ for $\kappa=3\times
10^{-3}$, $\lambda=10^{-4}$ (model A). The
white (not shaded) areas lead to the trivial
vacuum and are thus excluded. The upper
red/dark shaded area is excluded by the
requirement that, at $t_*$,
$V^{\prime\prime}\leq H_{\rm infl}^2$, while
the lower one corresponds to the quantum
regime. The blue/solid line shows the values
of $A_i$, $\phi_f$ which approximately
reproduce the correct value of the CMBR large
scale temperature anisotropy, as measured by
COBE (see Sec.~\ref{sec:toy} for details).}
\label{fig:model-4}
\end{figure}
\begin{figure}[tb]
\centering
\includegraphics[width=0.83\linewidth]
{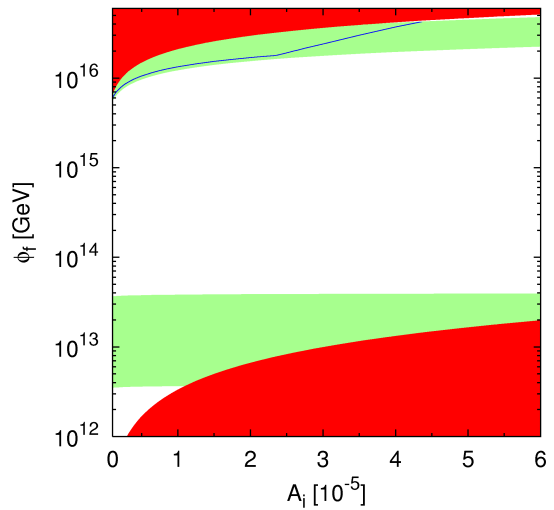}
\caption{The two green/lightly shaded bands
in the $A_i-\phi_f$ plane which lead to the
PQ vacua at $t_{\phi}$ for $\kappa=3\times
10^{-2}$, $\lambda=10^{-4}$ (model B). The
notation is the same as in
Fig.~\ref{fig:model-4}.}
\label{fig:model-5}
\end{figure}

It is clear that we must first impose the
requirement that $N_{\rm max}\geq N_Q$. The
time $t_*$ at which our present horizon scale
crosses outside the inflationary horizon can
then be determined from $N_Q=H_{\rm infl}
(t_f-t_*)$. Furthermore, we demand that, at
$t_*$, $V^{\prime\prime}\leq H_{\rm infl}^2$,
which ensures that this inequality holds for
all times between $t_*$ and $t_f$. This
condition, thus, guarantees that the PQ field
is effectively massless during the relevant
part of inflation and can act as curvaton. It
also provides us with an {\it a posteriori}
justification of the validity of the slow-roll
approximation used and ensures that the
velocity of $\phi$ at the end of inflation is
negligible. This masslessness requirement
yields an upper bound on $\phi_f$ for every
given $A_i$ and fixed values of $\kappa$ and
$\lambda$. The excluded region in the
$A_i-\phi_f$ plane for fixed $\kappa$ and
$\lambda$ is represented as a red/dark shaded
area in the upper part of this plane. In
Figs.~\ref{fig:model-4} and \ref{fig:model-5},
we show this upper red/dark shaded area for
$\kappa=3\times 10^{-3}$, $\lambda=10^{-4}$
(model A) and $\kappa=3\times 10^{-2}$,
$\lambda=10^{-4}$ (model B) respectively. The
lower red/dark shaded area in the $A_i-\phi_f$
plane corresponds to the quantum regime and is
also excluded for the reasons explained at the
end of Sec.~\ref{sec:curvature} (see
Figs.~\ref{fig:model-4} and \ref{fig:model-5}).
This also ensures that, at any given
instant of time during inflation, $\phi$ has
practically the same mean value in all comoving
volumes (which crossed outside the horizon).
So, during inflation, $\phi$ can be simply
considered as a function of time only.

\par
We start from any given value of $\phi_f$ at
$t_f$ which is not excluded by the above
considerations and assume vanishing time
derivative of $\phi$ at $t_f$. As explained
above, this is a good approximation provided
that $\phi_f$ does not belong to the upper
red/dark shaded area in the $A_i-\phi_f$
plane (see Figs.~\ref{fig:model-4} and
\ref{fig:model-5}). Moreover, we find
numerically that the subsequent evolution of
$\phi$ remains practically unchanged if we
take a small non-vanishing value of $\dot\phi$
at $t_f$. Under these initial conditions at
$t_f$, we follow the evolution of the field at
subsequent times ($t\geq t_f$) by solving the
classical equation of motion in
Eq.~(\ref{field-eqn}) with $H=2/(3t)$ for
$t_f\leq t\leq t_{\rm reh}$ and
$H=1/(2(t-t_{\rm reh}/4))$ for $t_{\rm reh}
\leq t\leq t_\phi$. In the latter expression
for $H$, we subtracted $t_{\rm reh}/4$ from
$t$ in order to achieve continuity of $H$ at
$t=t_{\rm reh}$. The time of reheating
$t_{\rm reh}$ is calculated by using
Eq.~(\ref{eq:reheat}) with $g_*=228.75$ and
the curvaton decay time $t_\phi$ is found by
employing Eq.~(\ref{Gphi}) with $\beta$ from
Eq.~(\ref{mu}), where $\mu$ is taken equal to
$300~{\rm GeV}$, which is smaller than half
the curvaton mass (see Eq.~(\ref{mphi})) so
that the decay of $\phi$ to Higgsinos is not
blocked kinematically. The parameter $\gamma$
in the effective PQ potential of
Eq.~(\ref{full-pot}) is taken equal to $0.1$
after the end of inflation (see
Sec.~\ref{sec:early}). We find that, for fixed
$\kappa$ and $\lambda$, there exist two bands
in the $A_i-\phi_f$ plane which lead to the
desired PQ vacua at $t_\phi$. They are
depicted as an upper and a lower green/lightly
shaded band (see Figs.~\ref{fig:model-4} and
\ref{fig:model-5}). The white (not shaded)
areas in the $A_i-\phi_f$ plane lead to the
false (trivial) minimum at $\phi=0$ and thus
must be excluded. Note that, for all relevant
$\kappa$'s and $\lambda$'s, the quantum
regime overlaps (at most) with the lower
green/lightly shaded band only. However, as
we will see later, this band is anyway
excluded by the data in all cases, the reason
being that it predicts an unacceptably large
isocurvature perturbation. So, the {\it a
priori} exclusion of the quantum regime does
not affect our results in any essential way.

\par
Besides models A and B (we summarize the
corresponding parameter values in
Table~\ref{tab:models}), which will be used as
our main examples in this presentation, we have
also studied three extra pairs of values of
$\kappa$ and $\lambda$, namely $\kappa=10^{-3}$
and $\lambda=10^{-4}$, $\kappa=10^{-2}$ and
$\lambda=10^{-4}$, and $\kappa=3\times 10^{-2}$
and $\lambda=3\times 10^{-4}$ (see below). In
the first two cases, the behavior was found to
be quite close to the behavior of model A as
depicted in Fig.~\ref{fig:model-4}, while the
latter case behaves similarly to the model B
(see Fig.~\ref{fig:model-5}).

\begin{table}
\setlength{\extrarowheight}{5pt} \centering
\begin{tabular}{|>{$}l<{$}c|c c|}
\hline
& Eq. & Model A & ~~~~~~~~~~~~~~~~Model B \\
\hline \hline
\kappa  & (\ref{W1}) & $3\times10^{-3}$ &
~~~~~~~~~~~~~~~~$3\times10^{-2}$ \\
\lambda & (\ref{W2}) &
\multicolumn{2}{c|}{$10^{-4}$}\\
m_{3/2}  & (\ref{PQ-pot}) &
\multicolumn{2}{c|}{$300$ GeV}\\
\vert A \vert & (\ref{PQ-pot}) &
\multicolumn{2}{c|}{5}\\
\mu  & (\ref{mu}) &
\multicolumn{2}{c|}{$300$ GeV}\\
\gamma & (\ref{eq:effec-mass}) &
\multicolumn{2}{c|}
{0 (0.l) during (after) inflation}\\
T_{\text{reh}} & (\ref{eq:reheat}) &
\multicolumn{2}{c|}{$10^9$ GeV}\\
m_{\text{LSP}} & (\ref{OmegaLSP}) &
\multicolumn{2}{c|}{$200$ GeV}\\
\hline
\end{tabular}
\caption{Summary of the fixed model parameters
for the two representative cases presented in
the text, called model A and B. The equation
where each parameter first appears is also
indicated.
\label{tab:models}}
\end{table}

\subsection{The calculation of $C_\ell$}
\label{sec:Cell}

\par
For any fixed pair of values for $\kappa$ and
$\lambda$, we take a grid of values of $A_i$
and $\phi_f$ which span the corresponding upper
or lower green/lightly shaded band. For each
point on this grid, we consider $\phi_f$ and
its perturbed value, which is found by adding
to $\phi_f$ the amplitude of $\delta\phi_f$
from Eq.~(\ref{dphif}). We then follow the
subsequent evolution of both these fields until
the time $t_\phi$ of the curvaton decay, where
we evaluate the amplitude of $\delta\rho_\phi/
\rho_\phi$. The amplitude $A_c$ of the partial
curvature perturbation from the curvaton is
then given by Eq.~(\ref{Ac}).

\par
We thus obtain a mapping of the values of
$\phi_f$ which are allowed for any given value
of $A_i$ onto the corresponding values of the
amplitude $A_c$ of the partial curvature
perturbation from the curvaton, which is the
relevant variable for the calculation of the
CMBR power spectrum. For the MC
analysis (see Sec.~\ref{sec:MC}), we
therefore use $A_i$ and $A_c$ as base
parameters describing the initial conditions
and limit our grid to values of $A_i$ smaller
than $6\times 10^{-5}$ (since larger values
would overpredict the temperature power of the
SW plateau). The spectral indices $n_i$ and
$n_c$ for each point of this grid are found by
using Eqs.~(\ref{epsiloni}) and (\ref{etai})
applied at $x=x_Q$, and Eqs.~(\ref{epsilonc}),
(\ref{etac}) and (\ref{nuc}) applied at $t_*$.
The fraction $f$ in Eq.~(\ref{eq:f}) is also
evaluated at $t_\phi$.

\par
The amplitude $A_a$ of the isocurvature
perturbation in the axions is calculated from
Eq.~(\ref{Aa}) with the initial misalignment
angle $\theta$ evaluated, for any given value
of the total CDM abundance $\Omega_{\rm CDM}h^2
=\Omega_{\rm LSP}h^2+\Omega_ah^2$, by using
Eqs.~(\ref{OmegaLSP}) and (\ref{Omegaa}) with
$m_{\rm LSP}=200~{\rm GeV}$. This value of
$m_{\rm LSP}$ corresponds \cite{wang}, in the
constrained MSSM with Yukawa unification
\cite{ananth}, roughly to the value of
$m_{3/2}$ ($=300~{\rm GeV}$) chosen here. Note
that, for our choice of parameters, the LSP
relic abundance is fixed ($\simeq 0.0074$). The
spectral index for axions in each point of the
grid is found by using Eq.~(\ref{nua}).

\par
In summary, for any fixed pair of values for
$\kappa$ and $\lambda$, we take a grid in the
variables $A_i$ and $\phi_f$ covering the
upper or lower green/lightly shaded band and
numerically map $\phi_f$ into the corresponding
value of the amplitude $A_c$ (which, of course,
depends on $A_i$ too). For any value of
$\Omega_Bh^2$ and $\Omega_ah^2$, we then
calculate the amplitudes squared of the
adiabatic and isocurvature perturbations using
Eqs.~(\ref{R2}), (\ref{Ric}) and
Eqs.~(\ref{S2}), (\ref{Sica}) respectively,
the cross correlation amplitude from
Eqs.~(\ref{C}), (\ref{Cic}) and the total CMBR
temperature and polarization power spectra via
Eqs.~(\ref{Cell})$-$(\ref{Cellcc}). The
curvaton fractional contribution to the
adiabatic amplitude $F^{\rm ad}_c$, the
dimensionless cross correlation $\cos\Delta$
and the entropy-to-adiabatic ratio $B$
are found from Eqs.~(\ref{Fc}) and
(\ref{cosDB}). For the MC analysis (see
Sec.~\ref{sec:MC}), we need to be able to
sample the $A_i-A_c$ space in any point as the
chains evolve. Therefore, we perform a
2-dimensional (2D) interpolation between the
points on the grid using a bicubic spline. A
little care is needed regarding the present
value of the Hubble parameter, which enters the
computation of the mass scale $M$ and thus
$H_{\rm infl}$ for given amplitude $A_i$ via
its impact on $N_Q$ (note that the first term
in the right hand side of Eq.~(\ref{NQ})
depends on $H_0$). As a consequence, the values
of the amplitudes $A_c$ and $A_a$ will also
depend on the value of $H_0$. However, we have
found numerically that changing the value of
$H_0$ around $H_0=72~\rm{km}~\rm{sec}^{-1}~
\rm{Mpc}^{-1}$ within the HST $1\sigma$ margin
(i.e. by $\pm 8~\rm{km}~\rm{sec}^{-1}~
\rm{Mpc}^{-1}$) has an impact less than about
$3\%$ on the computed quantities. Therefore, in
the computation of $M$ and of the amplitudes
$A_c$ and $A_a$, we fix the present Hubble
parameter to the HST central value $H_0=72~
\rm{km}~\rm{sec}^{-1}~\rm{Mpc}^{-1}$
\cite{hst}. Clearly, we do allow $H_0$ to vary
in the MC analysis (see below). In particular,
the dependence on $H_0$ of the amplitudes
$A_{{\rm P},i}$, $A_{{\rm P},c}$ and
$A_{{\rm P},a}$ at the pivot scale $k_{\rm P}$
(see Eq.~(\ref{AP})) is taken into account.

\subsection{A toy model for the CMBR}
\label{sec:toy}

\par
Before deploying the full MC machinery to
derive quantitative constraints on the
various parameters of our model, it is
instructive to consider a toy model for the
pre-WMAP CMBR data which allows us to
understand the salient qualitative features
of the parameter space of our initial
conditions. This approach has the great
advantage of offering a much more transparent
understanding of the parameter constraints
beyond the black box of the numerical MC study.

\par
A first rough idea about the viability of our
model can be obtained by using the approximate
expressions in Eqs.~(\ref{Cobe}) and
(\ref{fnell}) for the temperature SW plateau
and requiring that the COBE constraint in
Eq.~(\ref{ell10}) is fulfilled. To this end,
we take $\Omega_Bh^2=0.0224$ and
$\Omega_{\rm CDM}h^2=0.1126$, which are the
best-fit values from the WMAP measurements
\cite{wmap1}. (Note that, for this choice of
parameters, the axions constitute about $93\%$
of the CDM in the universe.) We find that the
COBE constraint cannot be satisfied in the
lower green/lightly shaded band in the
$A_i-\phi_f$ plane for any pair of values for
$\kappa$ and $\lambda$. The reason is
that the relatively low values of $\phi_f$ in
this band imply small values for $\phi_*$ too.
The amplitude $A_a$ then turns out to be quite
large as can easily be seen from
Eq.~(\ref{Aa}). This fact combined with the
sizable relic abundance of the axions leads to
large values of $S_a$ (see Eq.~(\ref{Sica})),
which yield an unacceptably large contribution
to the right hand side of Eq.~(\ref{Cobe}). In
the upper green/lightly shaded band in the
$A_i-\phi_f$ plane, on the contrary, the COBE
constraint is generally easily satisfied. The
resulting solution is depicted by a
blue/solid line (see Figs.~\ref{fig:model-4}
and \ref{fig:model-5}).

\par
We are further interested in deriving ball-park
estimates for the values of $A_i$, $A_c$. To
this goal, we again approximate the SW
temperature plateau by the expression in
Eq.~(\ref{Cobe}). As mentioned, this formula
neglects the effect of the cosmological
constant $\Lambda$ (late ISW term), but our
present purpose is to build an extremely
simplified toy model, not to include all
contributions which are fully taken into
account in the MC analysis below.
Therefore, we take one single datum, namely the
COBE constraint in Eq.~(\ref{ell10}), which
describes the height of the SW plateau, along
with its variance $\ell(\ell+1)\Delta
C^{\rm TT}_{\ell}/2\pi\big\vert_{\ell=10}
\approx 0.2\times 10^{-10}$ \cite{cobe10}.
Furthermore, we drop altogether the dependence
of Eq.~(\ref{Cobe}) on the spectral indices by
taking all perturbations to be scale invariant,
since the value of the indices predicted by our
model is anyway very close to unity (see
Sec.~\ref{sec:parcos}). We need a second datum
to be able to constrain two free parameters
($A_i$ and $A_c$) and this is given by the
height of the first adiabatic peak as measured
by BOOMERANG \cite{deBernardis}. In the
analysis of Ref.~\cite{Durrer}, this yields
$\ell(\ell+1)C^{\rm TT}_{\ell}/2\pi\big
\vert_{\ell= 212}\approx5.4\times 10^{-10}$
with error $\ell(\ell+1)\Delta
C^{\rm TT}_{\ell}/2\pi\big\vert_{\ell=212}
\approx 0.54\times 10^{-10}$. We model the
theoretical CMBR spectrum at the level of the
first adiabatic peak by retaining the adiabatic
contribution only, since the isocurvature
temperature spectrum drops very fast beyond the
SW plateau. Thus the prediction of our toy PQ
model for the height of the first peak is given
by
\begin{equation}
C_{212}^{\rm{TT,PQ}}\approx\frac{\Delta H}{25}
\left(R_i^2+R_c^2\right),
\end{equation}
where the constant factor $\Delta H\approx 6.5$
approximates the ratio of the first peak height
to the SW plateau for the adiabatic temperature
spectrum of the standard $\Lambda$CDM model.
Once again, this expression is very crude and
does not account for changes in $h$,
$\Omega_\Lambda$ ($\equiv\rho_\Lambda/\rho_c$
with $\rho_\Lambda$ being the dark energy
density), $\Omega_m$, $\Omega_B$ or $\tau_r$
(optical depth to the reionization epoch), all
of which affect the relative height of the peak
to the plateau, but it is sufficient for our
present goal. In summary, the likelihood
function $\like (A_i,A_c)$ of our toy model is
given by
\begin{eqnarray}
-2\ln\like (A_i,A_c)&\approx&\left(
\frac{C_{10}^{\rm{TT,PQ}}
(A_i,A_c)-C^{\rm{TT}}_{10}}
{\Delta C^{\rm{TT}}_{10}}\right)^2
\\
& &+\left(\frac{C_{212}^{\rm{TT,PQ}}(A_i,A_c)-
C^{\rm{TT}}_{212}}{\Delta C^{\rm{TT}}_{212}}
\right)^2,
\nonumber
\end{eqnarray}
where $C_{10}^{\rm{TT,PQ}}$ can be found from
Eq.~(\ref{Cobe}). The posterior (see
Appendix for definitions) is then
\begin{equation}
\pdf(A_i, A_c)\propto\like(A_i, A_c)\pi(A_i)
\pi(A_c).
\end{equation}
We adopt non-informative flat priors on the
amplitudes $A_i, A_c$, so that $\pi(A_i)=
\pi(A_c)=\text{constant}$. An alternative
choice is Jeffreys' prior, which is of the
form $\pi(A_i)=1/A_i$, $\pi(A_c)=1/A_c$
corresponding to a flat prior on $\log A_i$,
$\log A_c$. This prior implies that we do not
have any idea on the scale of $A_i$, $A_c$
before seeing the data and thus represents a
quite extreme choice of prior.

\begin{figure}[tb]
\centering
\includegraphics[width=\linewidth]{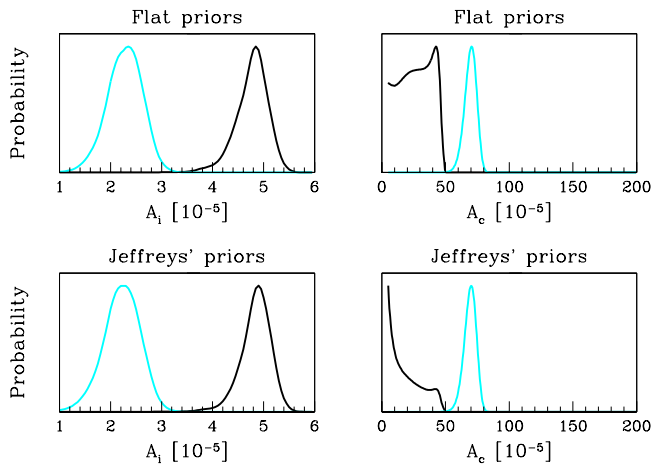}
\caption{Posterior marginalized pdf (normalized
at peak value) for the amplitudes $A_i$, $A_c$
from our toy model for two choices of priors:
flat priors (top panels), and Jeffreys' priors
(bottom panels). The black line is for model A
(upper green/lightly shaded band in
Fig.~\ref{fig:model-4}) and the cyan/light gray
line for model B (upper green/lightly shaded
band in Fig.~\ref{fig:model-5}). The
constraints on $A_i$, $A_c$ in model B and on
$A_i$ in model A as well as the upper limit on
$A_c$ in model A are robust with respect to the
choice of priors. Compare the top panels with
Fig.~\ref{fig:1dbase}, which shows the results
of the full MC analysis (with flat
non-informative priors).}
\label{fig:pqtoy}
\end{figure}

\begin{figure}[tb]
\centering
\includegraphics[width=\linewidth]
{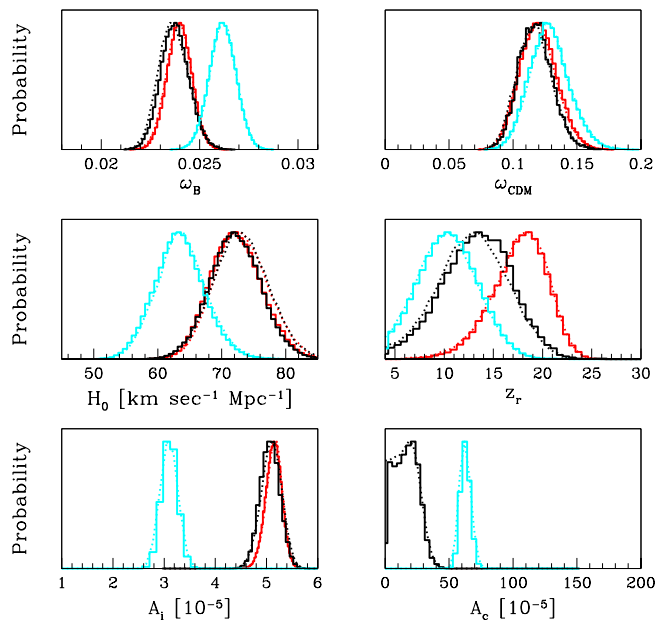}
\caption{1-dimensional (1D) marginalized
posterior distribution for the two cases
considered here: model A (black solid line) and
model B (cyan/light gray solid line) with only
the upper green/lightly shaded bands of
Figs.~\ref{fig:model-4} and \ref{fig:model-5}
included. The red/medium gray line is for the
standard pure inflaton single-field
inflationary case with Harrison-Zel'dovich (HZ)
spectrum ($n_i=1.00$ fixed), plotted here for
comparison. We plot non-smoothed histograms
corresponding to top-hat binnings to show the
resolution of our curves. Model A is quite
close to the pure inflaton case, except that the
curvaton contribution helps in reducing the
reionization redshift $z_r$. Model B displays a
preference for non-zero curvaton contribution,
a still lower $z_r$ and larger relic abundance
of baryons ($\omega_B=\Omega_Bh^2$) and CDM
($\omega_{\rm CDM}=\Omega_{\rm CDM}h^2$). Its
quality of fit is however poorer (see
Sec.~\ref{sec:modelcomp} for details). We also
display as dotted smoothed curves the values of
the mean posterior. All the curves are
normalized at their peak value.}
\label{fig:1dbase}
\end{figure}

\par
In Fig.~\ref{fig:pqtoy}, we present the
posterior marginalized probability distribution
functions (pdf's) for $A_i$, $A_c$ from our toy
model for the two choices of priors above. We
observe that there is a qualitative difference
between the results for the upper green/lightly
shaded band in Fig.~\ref{fig:model-4} and
\ref{fig:model-5} of model A and B. For model
B, both the inflaton and the curvaton amplitude
are well determined even in our over-simplified
toy model. From the plot, we read off that $A_i
\approx 2.3\times 10^{-5}$, $A_c\approx 70
\times 10^{-5}$. As a consequence, the choice
of prior hardly matters, as we would expect in
a situation where the posterior is dominated by
the likelihood. As for model A, we obtain that
$A_i\approx 4.9\times 10^{-5}$, but we can only
place upper limits on the value of $A_c$.
Since, in this case, $A_c$ is not a
well-determined parameter, the details of its
posterior pdf do depend on the prior.
Nevertheless, it is apparent that the upper
limit is robust, and we can deduce that $A_i
\lsim 50\times 10^{-5}$. These numbers have to
be taken only as ball-park estimates, and this
is why we do not bother to attach errors to
them. However, the comparison of
Fig.~\ref{fig:pqtoy} with
Fig.~\ref{fig:1dbase}, which shows the results
of the quantitative MC analysis including WMAP
and other recent CMBR data and all the other
cosmological parameters, displays an
astonishing agreement between the results of
the above toy model and of the full MC
analysis. The rather precise match of the two
results is actually the fortuitous outcome of
the cancellation between two opposite effects.
On the one hand, the MC analysis compared to
the toy model employs much more precise data,
which reduce the error on the constraints, but,
on the other hand, it also integrates over the
cosmological parameter space (totally ignored
in the toy model), which enlarges the error on
the marginalized quantities.

\par
From the study of the above toy model, we can
thus draw two conclusions. First, the height of
the first adiabatic peak to the large scale SW
plateau is the key quantity to constraining --
with at least order of magnitude accuracy --
the curvaton to inflaton contribution in the PQ
model. Clearly, quantitatively reliable bounds
need the inclusion of the detailed effect of
the cosmological parameters on the shape of the
power spectra (see next section). Second, we
have seen that the constraints on the inflaton
and curvaton amplitudes are robust with respect
to two different choices of (non-informative)
priors. For this reason, we will, from now on,
adopt a flat prior on $A_i$, $A_c$, which will
be used both for the extraction of the
constraints on the parameters (see
Sec.~\ref{sec:parcos}) and for the purpose of
Bayesian model comparison (see
Sec.~\ref{sec:modelcomp}).

\subsection{The Monte Carlo CMBR analysis}
\label{sec:MC}

\subsubsection{The setup}
\label{sec:setup}

\par
We now proceed to describe the setup of the
full numerical analysis confronting the
predictions of our PQ model with the CMBR
data. We constrain the relevant parameters of
our model by constructing Markov chains using
a modified version of the publicly available
Markov chain MC package \textsc{cosmomc}
\cite{cosmomc} as described in
Ref.~\cite{lewis}. As discussed in
Sec.~\ref{sec:spectrum}, the total CMBR angular
power spectrum is given by a (anti-)correlated
superposition of adiabatic and isocurvature
CMBR modes (see Eq.~(\ref{Cell})). The
adiabatic and isocurvature CMBR transfer
functions are computed in two successive calls
similarly to the technique employed in
Ref.~\cite{Valiviita}. For fixed values of
$\kappa$ and $\lambda$, the initial conditions
are completely specified by $A_i$ and $\phi_f$,
or equivalently by the values of $A_i$ and
$A_c$ as explained in Sec.~\ref{sec:Cell}. The
MC sampling takes as free parameters the
amplitudes $A_i$ and $A_c$, the physical baryon
and axion densities in the present universe
$\omega_B\equiv\Omega_Bh^2$ and
$\omega_a\equiv\Omega_ah^2$ in units of
$1.88\times 10^{-29}~{\rm g/cm}^3$, the
present value of the dimensionless Hubble
parameter $h=H_0/100\text{ km}\text{ sec}^{-1}
\text{ Mpc}^{-1}$, and the redshift $z_r$ at
which the reionization fraction is a half
(assuming sudden reionization). All other
derived quantities are computed from the above
parameter set, as detailed in
Sec.~\ref{sec:Cell}. In particular, the derived
parameters $R_i^2$, $R_c^2$, $F^{\rm ad}_c$,
$S_i^2$, $S_c^2$, $S_a^2$, $C_i$, $C_c$ are
obtained from Eqs.~(\ref{Ric}), (\ref{Fc}),
(\ref{Sica}) and (\ref{Cic}).

\par
For our choice of $m_{\text{LSP}}$ and
$T_{\rm reh}$, we have $\Omega_{\text{LSP}}h^2
\approx 0.0074=\text{constant}$, and the total
CDM abundance is $\omega_{\rm CDM}\equiv
\Omega_{\rm CDM}h^2=\Omega_{\text{LSP}}h^2+
\Omega_a h^2$. Our analysis considers flat
cosmologies only, thus the cosmological
constant energy density $\Omega_\Lambda$ (in
units of the critical energy density) is a
derived parameter, i.e. $\Omega_\Lambda=1-
(\omega_{\rm CDM}+\omega_B)/h^2$. We assume
three massless neutrino families and no
massive neutrinos (for constraints on these
quantities, see e.g. Ref.~\cite{constraints}).
We neglect the contribution of gravitational
waves to the spectrum, since the tensor to
scalar amplitude ratio at the SW
plateau is proportional to $\epsilon_i$,
which is completely negligible in our case.
In summary, for a fixed choice of $\kappa$
and $\lambda$, our parameter space is six
dimensional:
\begin{equation}
\params=\{\omega_B,\omega_a,h,z_r,A_i,A_c\}\;.
\end{equation}

\par
We compare the predicted CMBR temperature and
polarization power spectra to the WMAP
first-year data \cite{wmap1} (temperature and
polarization) with the routine for computing
the likelihood supplied by the WMAP team
\cite{Verde}. At a smaller angular scale, we
add the CBI \cite{cbi} and the decorrelated
ACBAR \cite{acbar1,acbar2} band powers as well.
We then run $N=20$ Markov chains starting from
randomly chosen points in the parameter space.
We take particular care in ensuring that the
starting points are spread over a wide range in
the $A_i-A_c$ plane. We then check that all
chains have converged to the same region of
parameter space. This indicates that this
region is indeed a global minimum (at least for
the range explored by the chains). This is the
main reason for using a relatively large number
of chains, since the danger that the chains are
stuck in a local minimum is great when
exploring mixed isocurvature initial conditions
(see e.g. Ref.~\cite{iso2}). A preliminary run
is needed to estimate the covariance matrix,
which is then diagonalized and used to perform
a final run until each chain contains
$M=30,000$ samples. The mixing of the chains is
checked using the Gelman and Rubin criterion
\cite{gelman}, for which we require that the
ratio of the variance of the mean to the mean
of the variance among the chains is $R<0.1$ for
all parameters. The parameter inference is
performed on the merged chains, which contain
around $5\times 10^5$ samples after the burn-in
has been discarded.

\par
As motivated above, we use flat top-hat priors
on the base parameters
\begin{equation}
\omega_B, \omega_a, z_r, A_i, A_c .
\end{equation}
The limits of the top-hat prior do not matter
for parameter estimation purposes, as long as
we check that the posterior density is
negligible near these limits. However, the
prior range of the accessible parameter space
plays an important role in computing the
Bayes factor for model testing (see
Sec.~\ref{sec:modelcomp}). We limit the maximum
range of the present dimensionless Hubble
parameter $h$ by imposing a top-hat prior $0.40
<h<1.00$ and we use the result of the HST
measurement \cite{hst}
\begin{equation}
\like^{\text{HST}}\propto\exp\left(\frac{h-h_0}
{2\sigma}\right)^2,
\end{equation}
where $h_0=0.72$ and $\sigma=0.08$, by
multiplying the likelihood function for the
CMBR data by the above Gaussian likelihood.

\par
Parameter constraints will be rather
insensitive to the details of the prior
distribution whenever the posterior is
dominated by the likelihood. As we have seen
from our toy model, the broad lines of the
constraints for the PQ model are indeed
robust with respect to the choice of
non-informative priors. We will see below that
the priors do play an important role in
Bayesian model comparison.

\begin{table}
\setlength{\extrarowheight}{5pt} \centering
\begin{tabular}{|>{$}l<{$}|>{$}c<{$}>{$}c<{$}|>
{$}c<{$}>{$}c<{$}|}
\hline
\text{param} & \multicolumn{2}{c|}
{\text{model A (upper band)}} &
\multicolumn{2}{c|}
{\text{model B (upper band)}}
\\
\hline
 & \text{best fit} & \text{1D }$68\%$
\text{ c.i.} & \text{best fit} &
\text{1D }$68\%$ \text{ c.i.}
\\
\hline\hline
\om_B & 0.024 & 0.024 \pm 0.002 & 0.026 & 0.026
\pm 0.002
\\
\om_a & 0.109 & 0.110 \pm 0.034 & 0.117 & 0.121
\pm 0.040
\\
H_0 & 72.4 & 72.2^{+11.5}_{-9.8} & 64.2 &
63.4^{+11.0}_{-9.2}
\\
z_r & 13.3 & 13.3^{+8.1}_{-9.3} & 10.4 &
10.5^{+7.7}_{-6.5}
\\
A_i\uf & 5.1 & 5.1 \pm 0.5 & 3.1 & 3.1\pm 0.4
\\
A_c\uf & 20.3 & <37.9\;(43.2) & 62.0 & 62.8\pm
10.4
\\
\hline
-\ln L_* & 721.2 & & 732.6 &
\\
\hline
\end{tabular}
\caption{Best-fit parameter values and sample
means with $1\sigma$ confidence intervals
(c.i.) for the 1D marginalized distribution
for the upper band of models A and B. We
indicate upper limits only when the parameter
is not constrained from below, in which case
the first number corresponds to the $68\%$ c.i.
(1 tail) and the number in parenthesis to the
$95\%$ c.i. of the parameter. The present
Hubble parameter $H_0$ is given in $\rm{km}~
\rm{sec}^{-1}~\rm{Mpc}^{-1}$. Finally, $L_*$
is the best-fit likelihood (see Appendix). For
comparison, the Harrison-Zel'dovich (HZ) pure
inflaton model has $-\ln L_*=721.7$.
\label{tab:bestfitbase}}
\end{table}

\subsubsection{Parameter constraints}
\label{sec:parcos}

\begin{figure}[tb]
\centering
\includegraphics[width=\linewidth]
{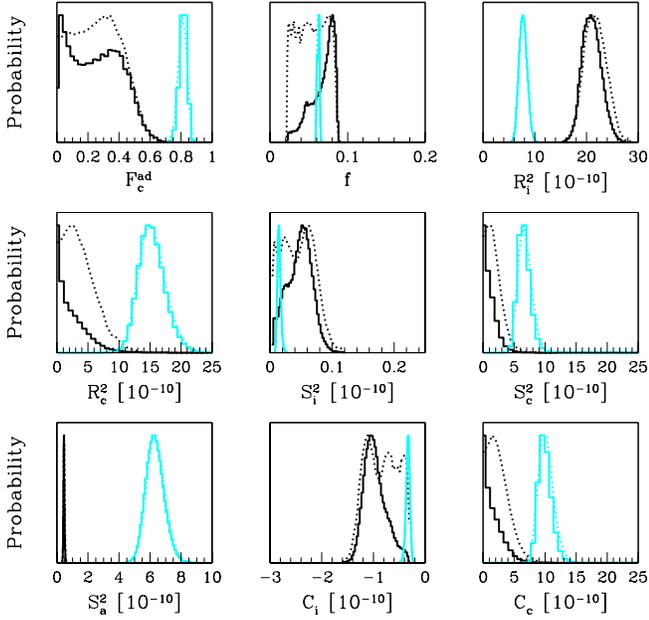}
\caption{As in Fig.~\ref{fig:1dbase}, but for
(most of) the derived parameters (the HZ pure
inflaton model is not included here). The
adiabatic amplitude in model B is dominated by
the curvaton ($F_c^{\rm ad}\approx 0.8$).}
\label{fig:1dderiv}
\end{figure}

\begin{figure}[tb]
\centering
\includegraphics[width=\linewidth]
{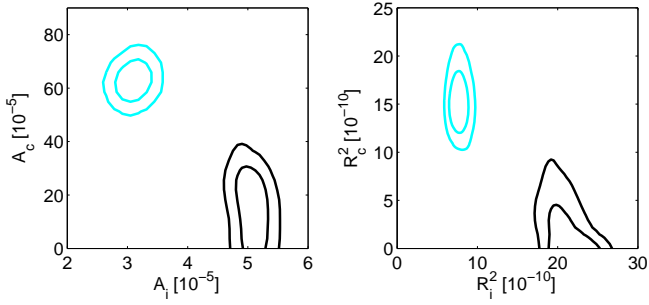}
\caption{Contours containing 68\% and 95\% of
the probability for the joint posterior pdf for
the upper band of models A and B (same
notation as in Fig.~\ref{fig:1dbase}). The
base primordial parameters $A_i$, $A_c$ are
displayed in the
left panel, while the right panel shows the
derived adiabatic amplitudes squared from the
inflaton ($R_i^2$) and the curvaton ($R_c^2$).
Model B prefers a non-vanishing curvaton
contribution to the adiabatic amplitude, but
its quality of fit is poorer (see text for
details).}
\label{fig:2dbase}
\end{figure}

In the Appendix, we summarize
some concepts and results from Bayesian
statistics which will be useful in the
following analysis (see e.g.
Refs.~\cite{MKbook} and \cite{bbook} for reviews
and details). We first consider the
upper green/lightly shaded band of models A and
B. In Tables~\ref{tab:bestfitbase} and
\ref{tab:bestfitderiv}, we present the best-fit
values and parameter constraints obtained from
the MC chains for our base and derived
parameters respectively. The 1-dimensional (1D)
marginalized posterior distributions for the
base and most of the derived parameters are
plotted, respectively, in
Figs.~\ref{fig:1dbase} and \ref{fig:1dderiv},
while the 2D contours of the posterior for
$A_i$, $A_c$ and the adiabatic amplitudes
squared $R_i^2$, $R_c^2$ are displayed in
Fig.~\ref{fig:2dbase}.

\begin{table}
\setlength{\extrarowheight}{5pt} \centering
\begin{tabular}{|>{$}l<{$}|>{$}c<{$}>{$}c<{$}|>
{$}c<{$}>{$}c<{$}|}
\hline
\text{param} & \multicolumn{2}{c|}
{\text{model A (upper band)}} &
\multicolumn{2}{c|}{\text{model B (upper band)}}
\\
\hline
 & \text{best fit} & \text{1D }$68\%$
\text{ c.i.} & \text{best fit} & \text{1D }
$68\%$\text{ c.i.}
\\
\hline \hline
R_i^2\ut & 20.4 & 21.0^{+6.2}_{-4.2} & 7.8 &
7.7 \pm 2.2
\\
R_c^2\ut & 2.5 & < 10.9\;(14.0)& 14.8 &
15.3^{+6.3}_{-4.9}
\\
S_i^2\ut & 0.06 & 0.05\pm0.04 & 0.015 &
0.015 \pm 0.006
\\
S_c^2\ut & 1.2 & < 5.0\;(7.1) & 6.6 &
6.6^{+3.2}_{-2.1}
\\
S_a^2\ut & 0.4 & 0.4 \pm 0.1 & 6.4 & 6.3
\pm 1.6
\\
C_i\ut & -1.1 & -1.0^{+0.6}_{-0.5}& -0.3
& -0.3 \pm 0.1
\\
C_c\ut & 1.8  & < 6.9\;(9.5) & 9.9 &
10.0^{+3.6}_{-2.3}
\\
\cos\Delta &0.08 & 0.0^{+0.5}_{-0.2} & 0.55 &
0.56 \pm 0.10
\\
B & 0.08 & <0.45\;(0.52) & 0.76 & 0.75 \pm 14
\\
\hline
F^{\ad}_c & 0.34 & < 0.60\;(0.67) & 0.81 & 0.81
\pm 0.07
\\
f  & 0.08 & 0.07^{+0.02}_{-0.05} & 0.06 & 0.062
\pm 0.004
\\
n_i & 0.988 &- & 0.982 & -
\\
n_c  & 1.011 & - & 1.003 & -
\\
n_a & 1.002 & - & 1.000 & -
\\
\hline
\end{tabular}
\caption{As in Table~\ref{tab:bestfitbase}, but
for the derived parameters. We do not give c.i.
for the spectral indices since the variation
in their value is less than $10^{-3}$.
\label{tab:bestfitderiv}}
\end{table}

\par
\begin{figure}[tb]
\centering
\includegraphics[width=\linewidth]
{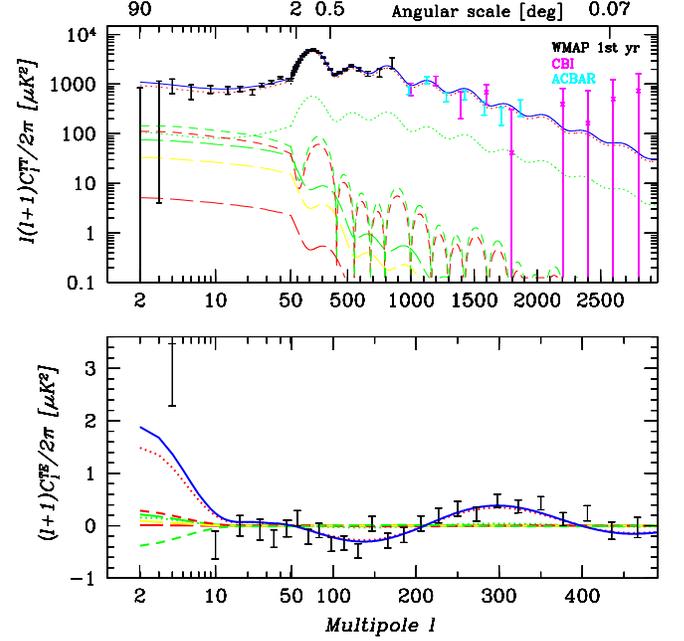}
\caption{Best-fit TT (upper panel) and TE
(lower panel) power spectra for the upper band
of model A (parameters as in
Table~\ref{tab:bestfitbase}). The line codes
are: red/dark gray for the inflaton,
green/medium gray for the curvaton, and
yellow/light gray for the axion contributions;
dotted for the adiabatic, long--dashed for the
isocurvature, and short--dashed for the
correlator parts. The total power corresponds
to the blue/solid line. In the upper panel, the
correlator parts are given in absolute value.
Note though that the inflaton correlator
contribution has to be added (negative
correlation), while the curvaton one has to be
subtracted (positive correlation) to obtain the
total power.}
\label{fig:BFpqMOD4}
\end{figure}

In the upper (green/lightly shaded) band of
model A, the adiabatic amplitude from the
inflaton dominates with only a modest
contribution ($\sim 10\%$ in amplitude squared)
to the adiabatic amplitude from the curvaton
and an even smaller curvaton isocurvature
amplitude and cross correlator. Note that the
axion isocurvature amplitude is negligible in
this case. Due to the small value of the
parameter $f$, the inflaton part of the
isocurvature amplitude squared (correlator) is
suppressed relative to the corresponding part
of the adiabatic amplitude squared by more than
two orders (one order) of magnitude. As a
consequence, the power
spectra are dominated by the adiabatic part of
the inflaton contribution. Moreover, on large
scales, we find that the sum of the adiabatic,
isocurvature and correlation parts of the total
power spectrum coming from the curvaton almost
cancels out, since the contribution from the
curvaton correlator is negative. We can easily
verify the behavior just described with the
help of Fig.~\ref{fig:BFpqMOD4}, where we plot
the best-fit power spectra for model A (upper
band) divided into the inflaton, curvaton and
axion contributions to their adiabatic,
isocurvature and correlator parts. The quality
of the fit, as expressed by the maximum
likelihood value $-\ln L_*=721.2$, is slightly
better than for
the pure inflaton Harrison-Zel'dovich (HZ)
case, which has $-\ln L_*=721.7$ (see, however,
our remarks below regarding model comparison).
This is not surprising since the curvaton
contribution turns out to play a modest role in
this model. In particular, the CMBR data put an
upper bound on the allowed value of the
curvaton amplitude, which is $A_c< 43.2\times
10^{-5}$ at $95\%$ confidence level (c.l.). The
total temperature power on large scales is
slightly larger than the pure adiabatic part --
the net effect is to increase the height of the
SW plateau compared to the height of the first
adiabatic peak. This mimics the impact of a
larger optical depth (and thus of a larger
$z_r$), and explains why model A shows a
preference for a later reionization than in the
pure inflaton case. The TE spectrum is
dominated by the inflaton adiabatic part, but
on large scales the curvaton and axion
contributions give a net power increase. This
effect helps in better fitting the
``reionization bump'' (i.e. the power increase
for $\ell\lsim 10$ due to reionization) at low
multipoles reducing the need for a rather
early reionization.

\begin{figure}[tb]
\centering
\includegraphics[width=\linewidth]
{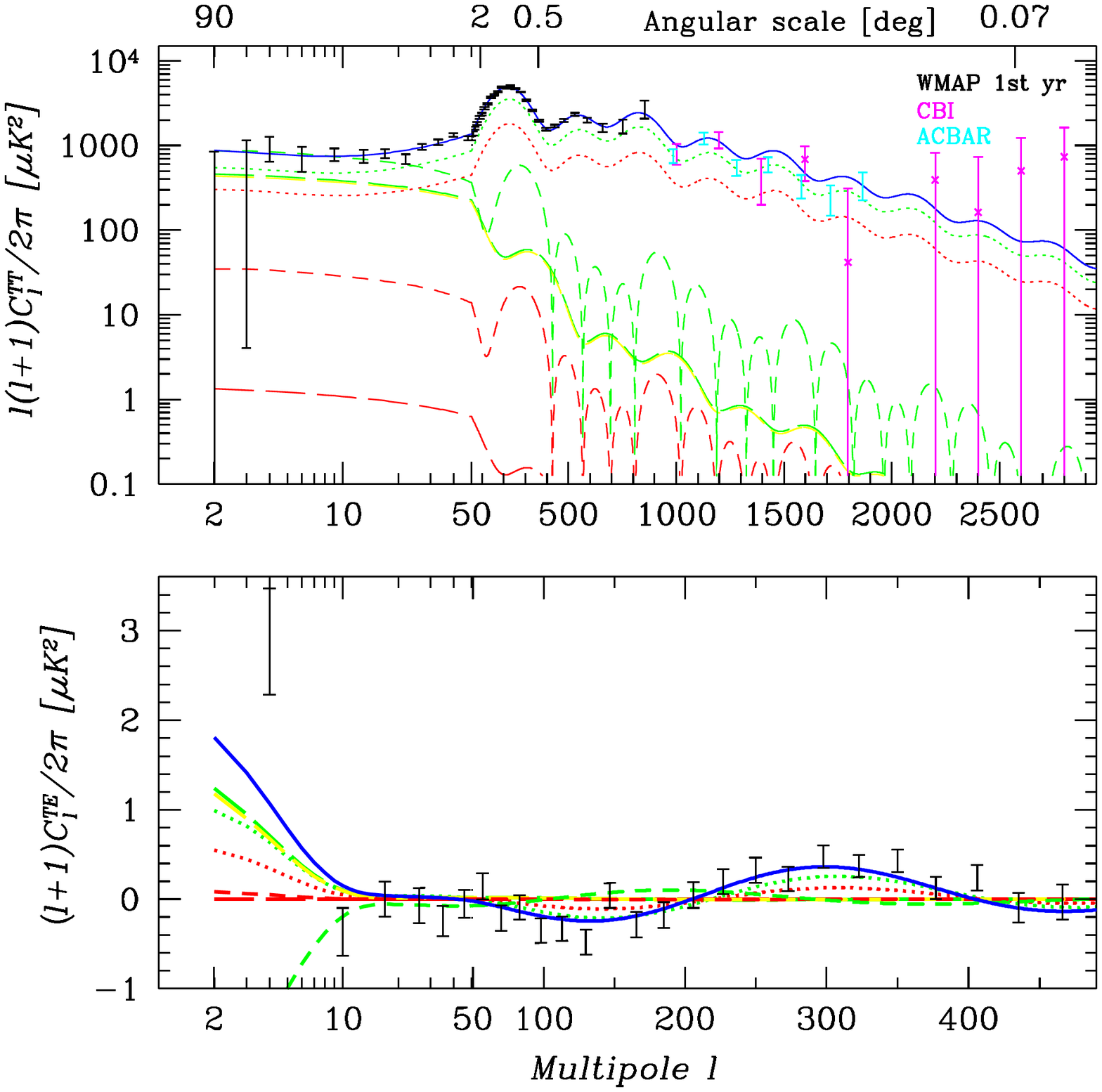}
\caption{Best-fit power spectra for the upper
band of model B. The base parameters are
according to Table~\ref{tab:bestfitbase}, and
the notation is as in Fig.~\ref{fig:BFpqMOD4}.}
\label{fig:BFpqMOD5}
\end{figure}

\begin{figure}[tb]
\centering
\includegraphics[width=\linewidth]{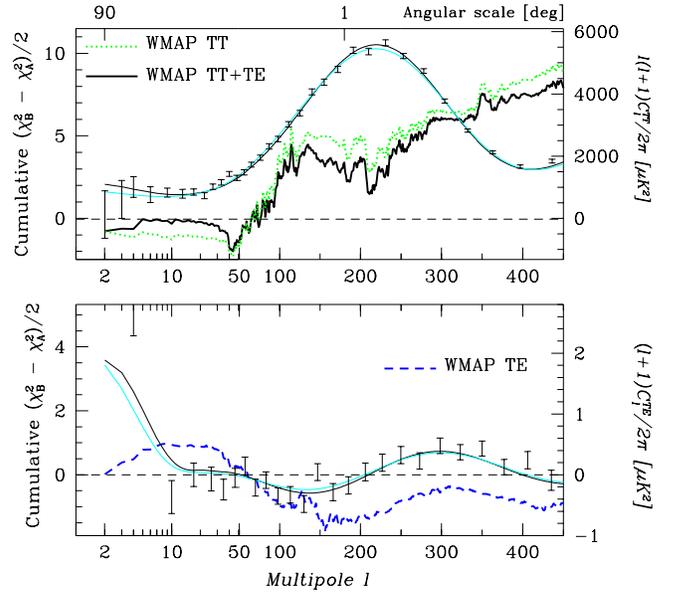}
\caption{Cumulative difference between model B
and model A (upper bands) of the quantity
$\chi^2/2\equiv -\ln L_*$ for the WMAP TT
(green/dotted line in the top panel) and TE
data (blue/dashed line in the bottom panel) in
units given on the left vertical axes. Their
sum is given in the top panel by the thick
black solid line. In the top panel, we
superimpose the two
corresponding best-fit TT spectra for model A
(thin black line) and model B (thin cyan/light
gray line) with units on the right vertical
axis. In the bottom panel, we superimpose the
two TE spectra (same notation). To compute the
$\chi^2$ difference as a function of the
multipole $\ell$, we use only the diagonal
elements of the data covariance matrix (for the
MC analysis, we, of course, included also the
off-diagonal elements, which however contribute
only a few percent). We also plot the binned
WMAP TT (top panel) and TE (bottom panel)
errorbars, as a guide to the eye to appreciate
the discriminative power of the WMAP data,
especially around the first acoustic peak in
the temperature spectrum. Model B is a better
fit to the TT SW plateau, since its power
there is smaller, but its TE spectrum in this
region fits the WMAP data worse. Model B cannot
reproduce the overall shape of the first
acoustic peak in temperature with enough accuracy,
and its goodness-of-fit is correspondingly
worse.}
\label{fig:chidiff}
\end{figure}

\begin{figure}[tb]
\centering
\includegraphics[width=\linewidth]
{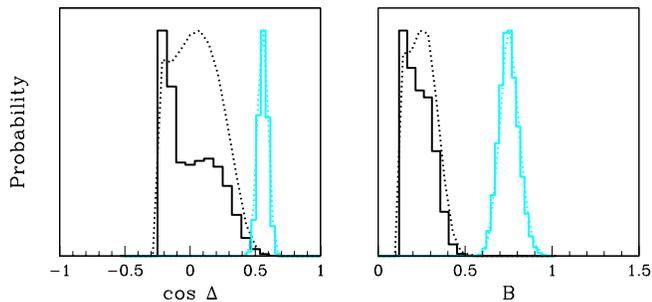}
\caption{As in Fig.~\ref{fig:1dbase}, but for
the dimensionless correlator $\cos\Delta$ and
the entropy-to-adiabatic ratio $B$ evaluated
at the pivot scale $k_P$ (see
Eq.~(\ref{cosDB})). (The HZ pure inflaton model
is not included.) The sharp drop for
$B\lsim 0.11$ encountered in model A is a
numerical feature due to the lower boundaries
of our MC run. The curve must accordingly be
interpreted as indicating an upper limit only,
which is given in
Table~\ref{tab:bestfitderiv}.}
\label{fig:1dcorr}
\end{figure}

\begin{figure}[tb]
\centering
\includegraphics[width=0.5\linewidth]
{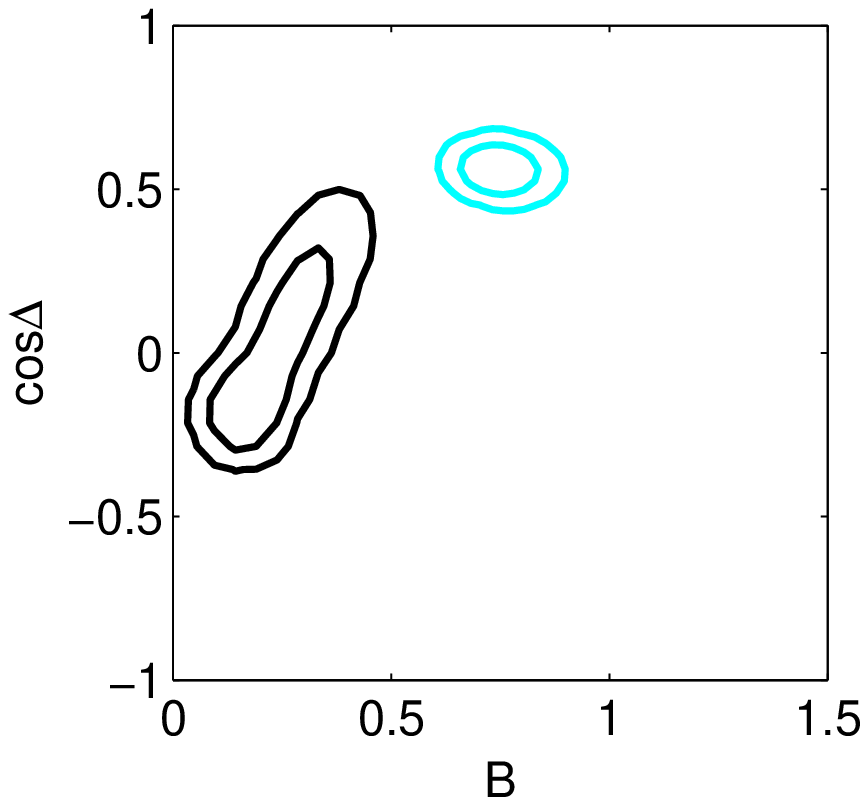}
\caption{Contours containing 68\% and 95\% of
the probability for the joint posterior pdf
for the upper band of models A and B for the
entropy-to-adiabatic ratio $B$ and
the dimensionless correlator $\cos\Delta$ (same
notation as in Fig.~\ref{fig:2dbase}).}
\label{fig:2dcomparecorr}
\end{figure}

\par
The upper (green/lightly shaded) band of model
B exhibits a preference for a non-vanishing
curvaton amplitude with very high significance
($A_c=62.8\pm 10.4$) as one can also see from
Fig.~\ref{fig:2dbase}. In this case, the
best-fit spectrum is a superposition of a
dominant curvaton adiabatic part and an
inflaton adiabatic contribution which is around
$50\%$ of the former in amplitude squared (see
Table~\ref{tab:bestfitderiv} and
Fig.~\ref{fig:BFpqMOD5}). This time, the
isocurvature curvaton part is sizable on large
scales, where however it is again cancelled by
the correlator. The large value of $\kappa$
yields a larger $H_*$ (see Fig.~\ref{fig:Hinf})
and thus pushes up the axion isocurvature
contribution (see Eq.~(\ref{Aa})),
which again adds power to the SW plateau. As in
model A (upper band), the isocurvature and
correlator inflaton parts are negligibly small.
For the best-fit parameters, the total power in
temperature in the COBE region is in better
agreement with the data, being slightly smaller
than for model A (upper band), thereby fitting
better the low-multipole region of the
spectrum. However, the overall quality of the
fit is worse ($-\ln L_*=732.6$) because the
model does not reproduce with enough precision
the shape of the first acoustic peak. We
illustrate this in Fig.~\ref{fig:chidiff},
where we compare the two best-fit spectra for
models A and B (upper bands). The WMAP data
around the first temperature peak are of such a
good quality that they can discriminate between
the two models thanks to the slightly different
shape of the peak. The reason why the
curvaton/inflaton mixture does not fit
accurately enough the first temperature peak in
model B is twofold. First, the isocurvature and
correlator contributions are still sizable in
the region of the first peak rise ($\ell\sim
100$), and this increases slightly the total
power in this part of the spectrum. Second, the
curvaton and inflaton have two slightly
different spectral indices (see
Table~\ref{tab:bestfitderiv}), and the resulting
tilts of the spectra are therefore mismatched.
As for the TE spectrum, the isocurvature axion
part plays an important role in reproducing
the reionization bump. Furthermore, model B
(upper band) shows a preference for a rather
high baryon abundance ($\omega_B\approx
0.026$), which is in strong tension with the
value indicated by BBN together with
observations of the light elements abundance,
which typically yields $\omega_B\approx 0.020
\pm 0.002$ \cite{Burles}.

\begin{table}
\setlength{\extrarowheight}{5pt} \centering
\begin{tabular}{|>{$}l<{$}|>{$}c<{$}|>{$}c
<{$}|}
\hline
\text{param} & {\text{model A (lower band)}} &
{\text{model B (lower band)}}
\\
\hline\hline
\om_B  & 0.029 & 0.080 \\
\om_a  & 0.171 & 0.010\star \\
H_0    & 50.1  & 100.0\star \\
z_r    & 19.4  & 30.0\star \\
A_i\uf & 0.10\star & 0.41 \\
A_c\uf & 75.0  & 300.0\star \\
\hline\hline
R_i^2\ut & 0.001 & 0.14\\
R_c^2\ut & 33.25 & 134.1\\
S_i^2\ut & \sim 10^{-4} & 0.002\\
S_c^2\ut & 9.05 & 971.4\\
S_a^2\ut & 14.37& 5512.1\\
C_i\ut & 0.003& 0.02\\
C_c\ut & 17.35& 360.9\\
\cos\Delta & 0.62 & 0.39\\
B &  0.84 & 6.95\\
\hline
F^{\ad}_c & 1.00 & 1.00\\
f         & 0.077& 0.39\\
n_i       & 0.986& 0.982\\
n_c       & 1.000&1.000\\
n_a       & 1.000 & 1.000\\
\hline
-\ln L_*  & 793.4 & 3014.6\\
\hline
\end{tabular}
\caption{Best-fit parameter values for the
lower band of models A and B. An asterisk
indicates that the corresponding parameter has
reached the limit of our top-hat prior in the
MC run, $H_0$ is again in
$\rm{km}~\rm{sec}^{-1}~\rm{Mpc}^{-1}$ and $L_*$
is the best-fit likelihood. Note that, due to
the large value of $\Omega_\Lambda
=0.91$ and the small value of $\Omega_m$ in the
best fit of model B (lower band),
the approximation used in deriving
Eq.~(\ref{Cobe}) is no longer valid.
\label{tab:bestfitbasefirstband}}
\end{table}

\begin{figure}[tb]
\centering
\includegraphics[width=\linewidth]
{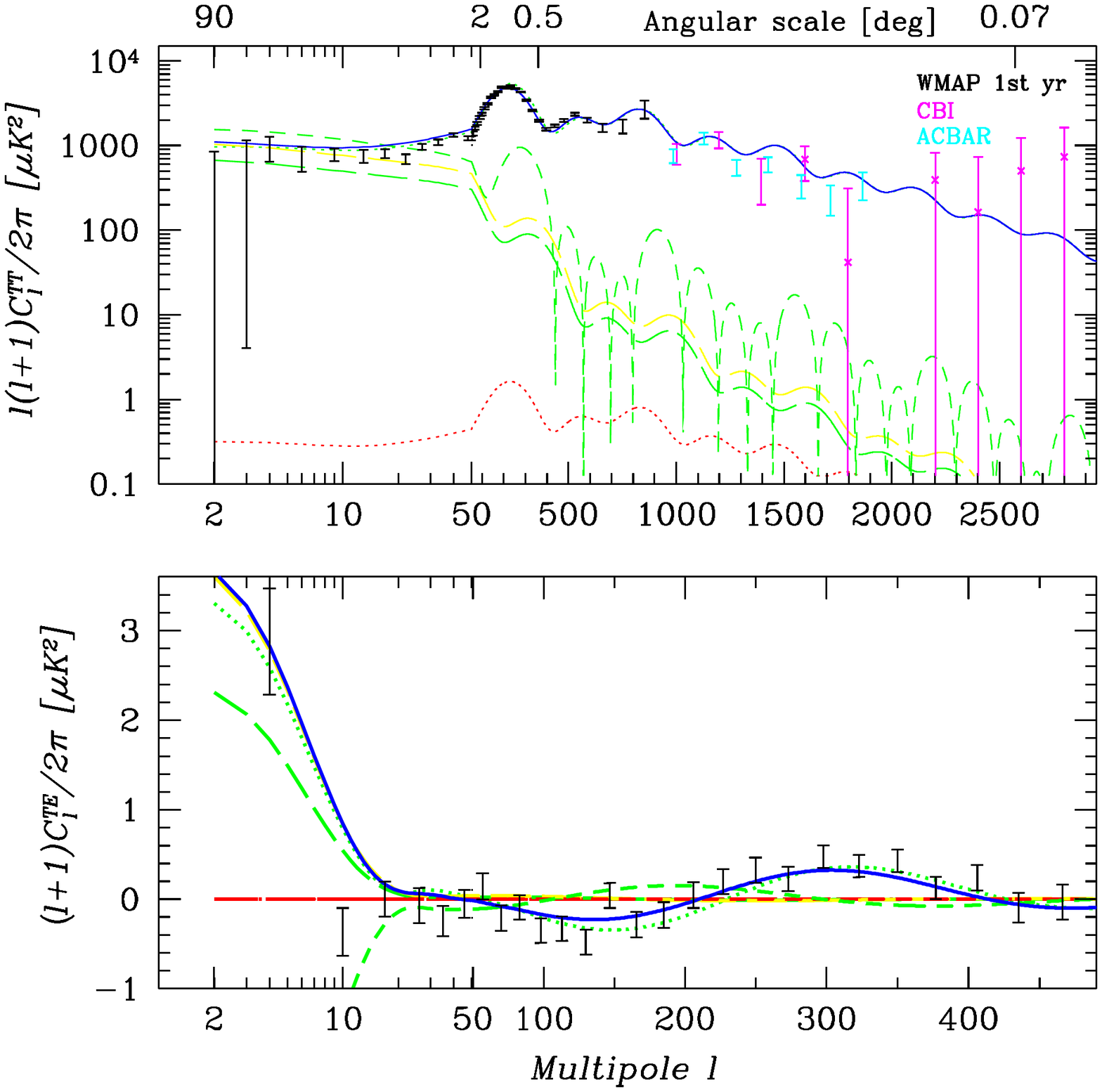}
\caption{Best-fit power spectra for the lower
band of model A. The base parameters are
according to
Table~\ref{tab:bestfitbasefirstband}, and
the notation is as in Fig.~\ref{fig:BFpqMOD4}.}
\label{fig:BFpqMOD4FST}
\end{figure}

\begin{figure}[tb]
\centering
\includegraphics[width=\linewidth]
{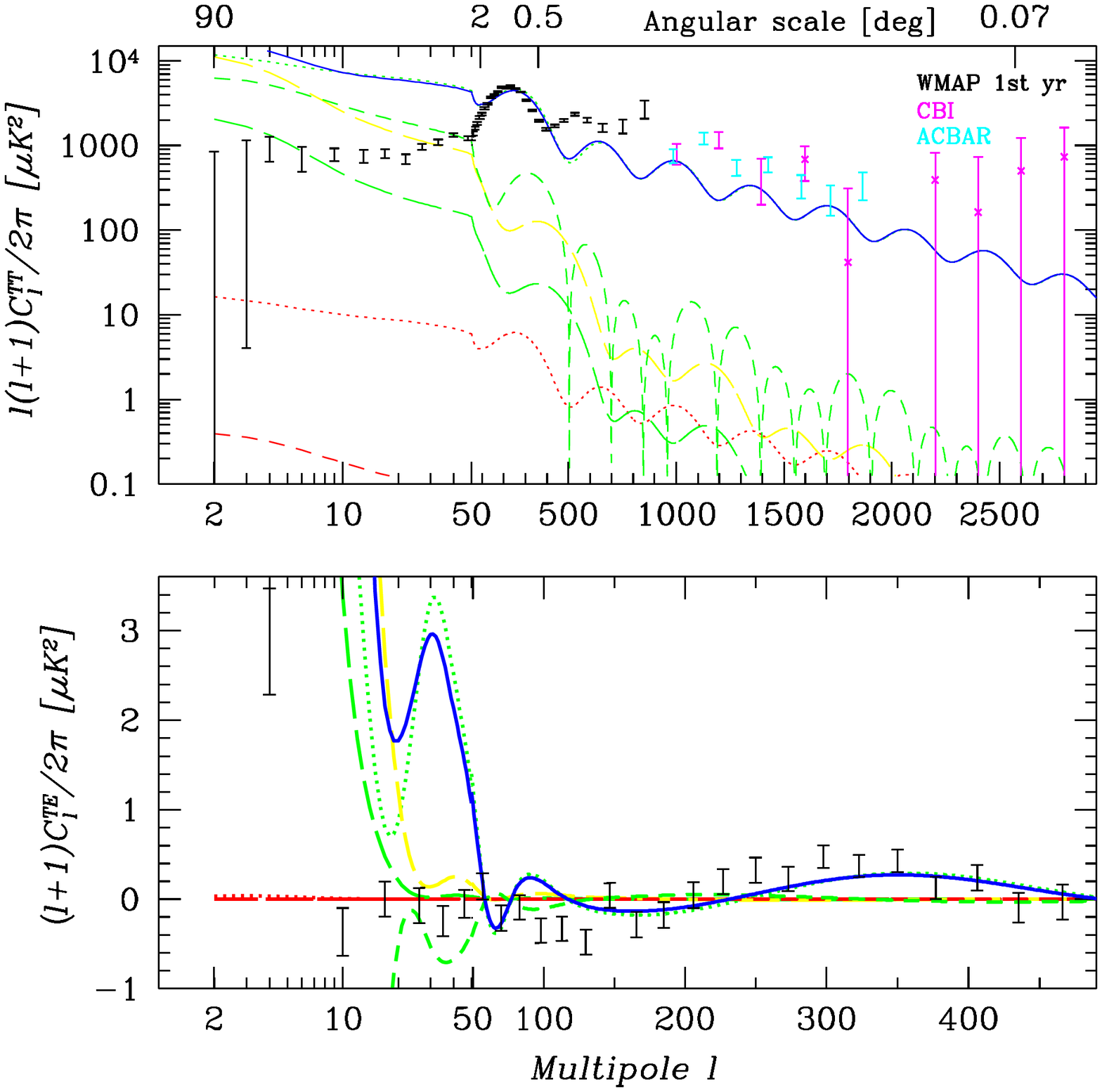}
\caption{Best-fit power spectra for the lower
band of model B. The base parameters are
according to
Table~\ref{tab:bestfitbasefirstband}, and
the notation is as in Fig.~\ref{fig:BFpqMOD4}.}
\label{fig:BFpqMOD5FST}
\end{figure}

\par
Fig.~\ref{fig:1dcorr} shows the 1D marginalized
constraints on the dimensionless correlator
$\cos\Delta$ and the entropy-to-adiabatic ratio
$B$, while Fig.~\ref{fig:2dcomparecorr} gives
the 2D joint constraints for these two
parameters. In model A (upper band), the
dominance of the inflaton and the low
amplitudes of the correlator modes result in a
value of the effective correlation $\cos\Delta$
roughly centered around zero. As discussed
above, the entropy contributions play a modest
role in model A (upper band), and
correspondingly we obtain the upper limit
$B\lsim 0.5$ (at $95\%$ c.l., 1 tail).
In model B (upper band), the large amplitude of
the curvaton correlator determines a positive
total correlation $\cos\Delta\approx 0.5$.
Together, the curvaton and axion isocurvature
amplitudes constitute a significant fraction of
the total amplitude, so that the
entropy-to-adiabatic ratio is much larger than
zero ($B\approx 0.75$). In summary, model B
(upper band) shows a complex superposition of
modes with a subtly balanced contribution of
adiabatic and isocurvature components.

\par
We have also performed a MC analysis for the
lower green/lightly shaded band in
Fig.~\ref{fig:model-4} of model A. As expected
from the arguments given above, the quality of
the best fit is poor ($-\ln L_*=793.4$) as one
can see from
Table~\ref{tab:bestfitbasefirstband} and
Fig.~\ref{fig:BFpqMOD4FST}, because
the small value of $\phi_*$ for this band gives
rise to a large axion contribution according
to Eqs.~(\ref{Aa}), (\ref{AP}) and
(\ref{Sica}). Furthermore, the curvaton
amplitude $A_c$ turns out to be much larger
than the amplitude $A_i$ from the inflaton.
This can be understood by observing that, in
contrast to $A_i$, the amplitude $A_c$ from the
curvaton can be large and still give a small
positive or even a negative contribution to the
total power in the SW plateau because the
curvaton correlation is positive and thus
subtracts power from large scales. This is
important since, in this case, the axion
contribution to the plateau is very large
leaving little room for other contributions,
while, at smaller scales, the axion
isocurvature spectrum becomes negligible and
the total temperature spectrum must necessarily
be dominated by either the curvaton or the
inflaton adiabatic contribution. So one of the
two amplitudes $A_i$ or $A_c$ has to be large
enough and, as we saw, this can happen only for
$A_c$. Therefore, at smaller scales, the
spectrum is dominated by the curvaton adiabatic
part. The best-fit inflaton amplitude $A_i$, on
the other hand, corresponds to the lower limit
of our parameter space as one can see from
Table~\ref{tab:bestfitbasefirstband}. The
reason is that the inflaton contribution to the
power spectra is always positive with an
unsuppressed (by factors of $f$ or
$(\Omega_{\rm LSP}+\Omega_B)/\Omega_m$)
adiabatic part. Thus the inflaton amplitude
$A_i$ must be very small in order not to
overpredict the large-scale power. Furthermore,
the TE spectrum has a very pronounced
reionization bump, which results from a rather
early reionization epoch and the large axion
isocurvature contribution.

\par
The lower band of model B is totally incapable
of producing a spectrum in reasonable agreement
with the data (see
Table~\ref{tab:bestfitbasefirstband}). For
values of $A_i$, $A_c$ corresponding to this
band, the axion contribution is huge and gives
temperature fluctuations at large angular
scales which are a few orders of magnitude
larger than what is observed (see
Fig.~\ref{fig:BFpqMOD5FST}).

\par
Regarding the issue of non-Gaussianity, we note
that our parameter $f$, defined in
Eq.~(\ref{eq:f}), is $\gg 10^{-2}$. Therefore,
non-Gaussianity from the curvaton partial
curvature perturbation is well within the
current bounds from WMAP \cite{wmapgauss} (see
Ref.~\cite{curv3}). Actually, the values of $f$
in our model are high enough to ensure that
this statement remains true even in the limit
of pure curvaton. Moreover, the
non-Gaussianity of the isocurvature
perturbation in the axions is also negligible.
This is because the perturbation
$\delta\theta=H_*/2\pi\phi_*$ acquired during
inflation by the initial misalignment angle
$\theta$ (see e.g. Ref.~\cite{dllr1}) is always
much smaller than $\theta$. As a consequence,
terms which are second order in this
perturbation can be safely neglected (see
Ref.~\cite{axionnongauss}). Finally, the
non-Gaussian component from the inflaton is
also negligibly small.

\par
We have performed two other MC runs with the
same value for $\lambda=10^{-4}$ as in model A
and $\kappa$ slightly larger or smaller, i.e.
$\kappa=10^{-2}$ or $\kappa=10^{-3}$,
recovering a behavior which is qualitatively
similar to the behavior of model A. We have
thus chosen to present our results for this
particular value of $\kappa$ ($=3\times
10^{-3}$) as representative of the behavior of
this class of models. We have also tried a
slightly larger value of $\lambda$
($=3\times 10^{-4}$) for the same value of
$\kappa=3\times 10^{-2}$ as in model B, and
found a behavior qualitatively similar to
model B.

\subsubsection{Bayesian model comparison}
\label{sec:modelcomp}

\par
So far we have been concerned with the
problem of deriving parameter constraints from
the data, given an underlying model for the
generation of the primordial fluctuations.
Models A and B (upper bands) actually both
belong to a continuous class of models
(belonging to our PQ model) which is
parameterized by $\kappa$ and $\lambda$.
However, since in this work we have chosen to
fix the values of $\kappa$ and $\lambda$, we
may as well consider models A and B (upper
bands) as two discrete disconnected models.
The question is then to compare these two
models with the standard pure inflaton HZ model
and decide which of these three models is most
favored by data. It should be stressed that
model comparison (or model testing) is a
different issue than parameter extraction, and
indeed it represents a further step in Bayesian
inference. In fact, it can very well be that
model testing arrives at a different conclusion
than parameter estimation. Indeed, it can be
that the estimated value of a parameter under
a model $\mdl_1$ is far from the null value
predicted by model $\mdl_2$, but $\mdl_1$ is
disfavored against $\mdl_2$ by Bayesian model
testing, a fact sometimes called ``Bartlett's
paradox'' \cite{Lindley}. This is exactly the
case for $A_c$ in model B (upper band) compared
to the pure inflaton model with flat spectrum
of perturbations, as we will show below.

\par
Sampling statistics (i.e. the frequentist
approach to parameter estimation) uses the
notion of the goodness-of-fit test to
assess the viability of a model without the
need of specifying an alternative hypothesis.
This usually reduces to the $\chi^2$ statistics
for the observed data, presuming the model
under consideration, $\mdl$, is true.
$\mdl$ is then rejected if the value of the
goodness-of-fit falls above a certain
threshold in the tail of the distribution. If
we take this criterion at face value and use
the $\chi^2$ statistics for the WMAP data, the
best-fit pure inflaton $\Lambda$CDM model with
$n_i\neq 1$ having $\chi^2=1431$ for $\nu=1342$
degrees of freedom (see last paper in
Ref.~\cite{wmap1}) would be rejected with a
type I error (i.e. probability of falsely
rejecting a true model) of about $5\%$. Notice
that this does not mean, as sometimes stated,
that ``the probability of the model is $5\%$''.
We will see below that Bayesian model
comparison is more informative and can give
useful guidance for model building.

\par
It is clear that models with a very poor best
fit can be discarded simply by inspection. An
extreme example is certainly the lower band of
model B presented above. The goodness-of-fit
for the lower band of model A is also
very poor, even though this is not readily
distinguishable from Fig.~\ref{fig:BFpqMOD4FST}
due to the logarithmic scale. In fact, the
best-fit point has $-\ln L_*=793.4$, compared
to $-\ln L_*=721.2$ for the upper band of this
model. Again, we can dismiss the lower band
without further analysis. However, we need a
more powerful tool if we are to decide, on the
basis of the available data, between the
viability of our PQ model as opposed to the
pure inflaton HZ model. Bayesian inference
offers a natural tool for model comparison in
the form of the evidence in favor of the model
(see Appendix for details and precise
definitions).

\begin{figure}[tb]
\centering
\includegraphics[width=\linewidth]
{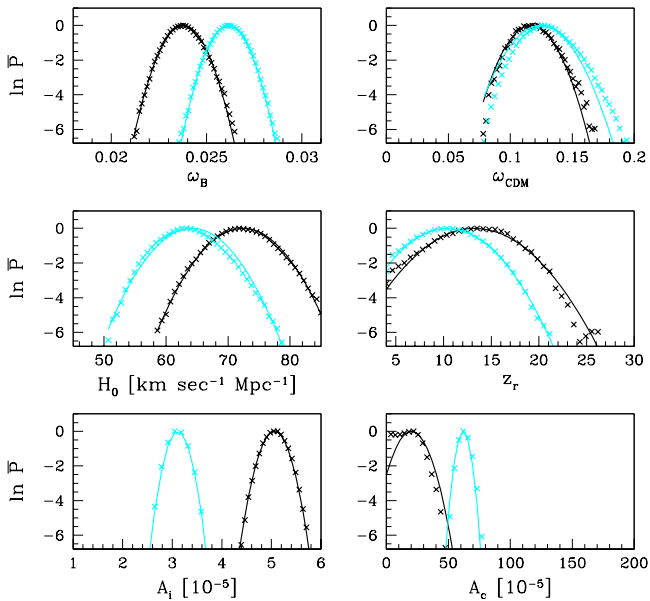}
\caption{Illustration of the Laplace
approximation to the (non-normalized) posterior
used to compute the model evidence. For our
base parameters, we plot the logarithm of the
1D marginalized posterior from the MC samples
(crosses corresponding to the histograms in
Fig.~\ref{fig:1dbase}) along with the
corresponding Laplace approximation (solid
smooth lines) of Eq.~(\ref{eq:laplace}) for the
upper bands of model A (black) and model B
(cyan/light gray). Clearly, the Laplace
approximation is very good for model B, and of
acceptable quality for model A. All the curves
are normalized to zero at their peak value.
In particular, $A_c$ is a rather non-Gaussian
direction for model A (upper band), since the
posterior from the MC samples only gives an
upper limit on this parameter.}
\label{fig:laplace}
\end{figure}

\par
We thus compute the Bayes factor for models A
and B (upper bands), comparing each of them to
the scale-invariant (i.e. $n_i = 1.00$) pure
inflaton model, which we call in the following
the HZ inflation model. We approximate the
integrals involved in the evaluation of the
Bayes factor by using the Laplace approximation
(see Eq.~(\ref{eq:laplace})). We must first
check that a multi-dimensional Gaussian is a
reasonable approximation to the actual shape of
our (non-normalized) posterior pdf. This is
shown in Fig.~\ref{fig:laplace}, where we plot
the logarithm of the 1D marginalized
non-normalized posterior from the MC samples
along with the corresponding
Laplace approximation. For model B (upper band),
the Gaussian approximation is quite accurate
along all directions, while for model
A (upper band) we notice that $A_c$ is a rather
non-Gaussian direction, which is not surprising
since this model only gives upper bounds on
the amplitude $A_c$. However, inspection of the
figure does seem to support the view that it is
reasonably accurate to use the Laplace
approximation to compute the model evidence.
Another qualitative criterion is the similarity
of the marginalized 1D posterior and the mean
posterior. If the posterior is a
multi-dimensional Gaussian, then the two curves
coincide (see e.g. Ref.~\cite{lewis}).
From Fig.~\ref{fig:1dbase}, we can indeed
confirm that the two curves are very similar,
again strengthening the conclusion that the
Laplace approximation is legitimate in our
case, more so for model B.

\par
We first compare the upper band of model B
($\mdl_1$ in the notation used in the
Appendix) against the HZ inflationary case
($\mdl_2$), and compute the
Bayes factor $B_{12}$ in favor of model B
(upper band). The term $\mathcal{L}_{12}$ (see
Eq.~(\ref{eq:L12})) is just the difference of
the best-fit log-likelihoods yielding
$\mathcal{L}_{12}=-10.9$, and clearly disfavors
model B (upper band), whose fit is worse. The
term $\mathcal{C}_{12}$ (see Eq.~(\ref{eq:C12}))
describes the ratio of the occupied volumes in
parameter space by the posterior pdf's of the
two models times a factor taking into account
the different dimensionality of the two
parameter spaces and is found to be
$\mathcal{C}_{12}=3.0$. Further considerations
are needed to evaluate $\mathcal{F}_{12}$ (see
Eq.~(\ref{eq:F12})), the term which reflects
the prior available volume of parameter space
under each model. In the present case, we do
not need to specify the top-hat range of the
priors on the parameters which are common to
both model B (upper band) and the HZ inflation
model. This is because, whatever prior belief
we hold about the possible range for these
parameters, it will be the same for both
models. The parameters common to both models
are $\omega_B$, $\omega_{\rm CDM}$, $h$, $z_r$
and $A_i$. In the PQ model, the prior range for
the CDM abundance actually applies to the axion
component of CDM only, but the difference,
which is caused by the presence in the universe
of the LSPs, is very small and insignificant
for our considerations here. Also, we could
actually assign to the inflaton amplitude $A_i$
two different prior ranges for the PQ and the
HZ inflation models if we had any reason to
believe that the ranges can be significantly
different in the two cases. Here we take the
view that $A_i$ is essentially a free parameter
in both models, and thus whatever prior range
we assign to it will cancel out from
$\mathcal{F}_{12}$. As a consequence, the only
prior range which does not cancel out from
$\mathcal{F}_{12}$ is the one for the extra
parameter of our PQ model, i.e. $A_c$. So we
have
\begin{equation}
\mathcal{F}_{12}=\ln\frac{1}{\Delta A_c},
\end{equation}
where $\Delta A_c$ is the top-hat prior range
of $A_c$ for the PQ model which is allowed in
model B (upper band). The bounds of the prior
on $A_c$ correspond to the limits on $A_c$ in
the upper band of model B. The lower limit is
of no importance, since we can simply set it
equal to zero. The upper limit is achieved on
the boundary of the upper band, which gives
$A_c^{\rm{max}}\approx 0.1$. From these
considerations and writing $A_c$ in units of
$10^{-5}$ (which are the same units used in
the covariance matrix), we have $\Delta A_c
\approx 10^4$, and thus $\mathcal{F}_{12}=
-9.2$. Putting everything together, we find
$\ln B_{12}=-17.1$ and thus the Bayes factor
disfavors model B (upper band) against the HZ
inflation model with odds of about $10^7$
against $1$. Notice that the
reason for such high odds comes partly from the
worse quality of fit of model B (upper band),
and partly from an ``Occam's razor'' penalty of
the PQ model, because it introduces a new scale
in the problem (the curvaton amplitude $A_c$)
which has a very wide prior range. We comment
further on this aspect below.

\begin{table}
\setlength{\extrarowheight}{5pt}\centering
\begin{tabular}{|>{$}l<{$}|>{$}c<{$}|>{$}c<{$}|
}
\hline
 & {\text{model A}} &
{\text{model B}}
\\
& {\text{(upper band)}} &
{\text{(upper band)}}
\\
 & {\text{versus HZ}} & {\text{versus HZ}}
\\
\hline \hline
\mathcal{L}_{12}
& 0.5 & -10.9
\\
\mathcal{C}_{12}
& 4.8 & 3.0
\\
\mathcal{F}_{12}
& -9.2& -9.2
\\
\hline
\ln B_{12}
& -3.9& -17.1
\\
\text{posterior odds}
& 1:50& 1:10^7
\\
\hline
\end{tabular}
\caption{Results of the Bayes factors analysis
for comparing the PQ models A and B (upper
bands) to the HZ inflation model. When
considering CMBR data only, the odds are
in favor of the HZ model against the PQ models.
In particular, model B (upper band) is very
strongly disfavored.
\label{tab:evidence}}
\end{table}

\par
We now compute the Bayes factor for the upper
band of model A ($\mdl_1$) against the HZ
inflation model ($\mdl_2$). As mentioned above,
in this case we expect the Laplace
approximation to be less accurate, because
$A_c$ is now a non-Gaussian direction. The term
$\mathcal{L}_{12}$ is now slightly in favor of
model A (upper band), since its fit is
marginally better than the one of the HZ
inflation model, giving $\mathcal{L}_{12}=0.5$.
From the covariance matrices of the two models,
we obtain $\mathcal{C}_{12}=4.8$, while the
same reasoning as above gives again a very
small value for $\mathcal{F}_{12}$ as a
consequence of the large allowed prior range of
$A_c$, i.e. $\mathcal{F}_{12}=-9.2$. This last
term again heavily penalizes the PQ model,
resulting in $\ln B_{12}=-3.9$, or odds of
$50:1$ against the PQ model A (upper band). The
Laplace approximation is however likely to
underestimate the actual volume occupied by the
likelihood function in parameter space, due to
the fact that $A_c$ is a rather non-Gaussian
direction for model A. As a consequence,
the above odds can be interpreted as an upper
limit for the evidence against model A. We
summarize these results on the PQ models A and
B (upper bands) in Table~\ref{tab:evidence}.

\par
One must be very careful when interpreting the
above results for the evidence. In fact, the HZ
inflationary model, which we used for the
comparison, is a natural benchmark model as far
as the CMBR data are concerned. However, one
must keep in mind that the PQ model which we
describe in this work addresses many
fundamental issues which lie outside the scope
of the phenomenological HZ inflation model,
such as the strong CP and $\mu$ problems of
MSSM, the generation of the observed BAU and
the nature of the CDM in the universe. Our
approach was to consider fundamental (SUSY GUT)
models of particle physics within the
cosmological framework by merging together
requirements and constraints both from the
particle physics and the cosmology side. In
general, it is clear that any viable particle
physics model has many free parameters about
which little or no experimental information is
available at the moment, such as the parameters
of the electroweak Higgs sector, the sparticle
mass spectrum or the boundary conditions from
supergravity which also determine the radiative
electroweak symmetry breaking. Nevertheless, we
do have some indication about their order of
magnitude by applying criteria of simplicity,
naturalness and elegance to our fundamental
theories. At this stage, the exquisite
cosmological data nowadays at our disposal can
provide significant constraints on the
parameter space of the model. Our evidence
calculation, in fact, takes into account only a
part of our knowledge (i.e. the CMBR data)
neglecting all the other issues which are
addressed at a fundamental level by our model.
It seems fair to say that, on the whole (i.e.
considering both cosmology and particle
physics), the PQ model presented in this work
aims at a broader explanatory power than a
phenomenological inflation model. We thus
conclude from our analysis that the present
CMBR data can strongly disfavor certain regions
of parameter space, as it is the case for model
B. In this case, we found robust evidence that
a mixture of curvaton and inflaton
contributions to the cosmological perturbations
is in disagreement with the present CMBR data.
On the other hand, the odds against model A
should be regarded while keeping in mind the
above considerations. In conclusion, it seems
to us that, at the present stage, we cannot
reject the possibility of a subdominant curvaton
contribution to predominantly inflaton-dominated
adiabatic perturbations.

\par
On a more phenomenological level, our treatment
of the evidence highlights a generic feature of
any model which introduces a second (or
several) scale-free parameter(s) to describe an
extra non-adiabatic component, namely that
Occam's razor of Bayesian model comparison will
always penalize such a model with respect to
the pure inflaton single-field HZ inflationary
model by assigning to it a very small
$\mathcal{F}_{12}$ term. This reflects the fact
that the extra amplitude parameter
(describing an isocurvature contribution) can
{\it a priori} assume any value at all, and
therefore there is no hard-wired justification
why its value should be smaller than $10^{-5}$,
or indeed close to but different from zero. So,
from a Bayesian point of view, it is certainly
unjustified to increase the model's complexity
to achieve only a minuscule gain (if any at
all) on the quality of fit. Traditionally, most
attention has been devoted to the maximum
likelihood value as the criterion to judge
whether a certain new parameter is useful or
not. But, from the point of view of model
building and Bayesian model testing,
restricting the prior volume of parameter space
could end up to be as useful, for an
inflationary model which would predict the
value of $A_i$ to be in the ball-park of
$10^{-5}$ would have very favorable posterior
odds against a model in which $A_i$ is
essentially a free parameter.

\par
Finally, the example of model B (upper band)
above strikingly illustrates that Bayesian
``credible intervals'' cannot be interpreted as
``significance levels'', for, just by looking
at the constraints on the amplitude $A_c$ for
the upper band of model B, one could have
(erroneously) deduced a $6\sigma$ detection of
a non-zero curvaton amplitude. Clearly, from
the poor quality of the best fit
($-\ln L_*=732.6$), it would have been
immediately obvious, without the need of
computing the Bayes factor, that this
particular model could not be favored by data.
However, the conceptual point remains: the
question whether a certain parameter value
differs from the null reference value cannot be
answered by looking at the confidence intervals
(c.i.), but must be tackled by proper model
comparison procedures. While this point is
certainly clear to data analysts working in the
field, it seems appropriate to stress it once
more in view of a widespread misinterpretation
of this concept.

\section{Conclusions}
\label{sec:conclusion}

\par
We considered a simple and concrete SUSY GUT
model which automatically and naturally solves
the strong CP and $\mu$ problems of MSSM via a
PQ and a continuous R symmetry. This model
also leads to the standard SUSY realization of
hybrid inflation. The PQ field of this model,
which corresponds to a flat direction in field
space lifted by non-renormalizable interactions,
can act as curvaton contributing together with
the inflaton to the total curvature
perturbation in the universe. In contrast to
the standard curvaton hypothesis, we did not
suppress the contribution from the inflaton.

\par
The CDM in the universe predicted by this model
consists predominantly of axions which are
produced at the QCD phase transition. It also
contains LSPs originating from the gravitinos
which were thermally produced at reheating and
decayed well after BBN. For simplicity, we
assumed that
there are no thermally produced LSPs in the
model. The baryons, which are generated via a
primordial leptogenesis occurring at reheating,
as well as the LSPs inherit the partial
curvature perturbation of the inflaton. This is
different from the total curvature perturbation,
which receives a contribution from the curvaton
too. Therefore, the baryons and LSPs carry also
an isocurvature perturbation, whose correlation
with the total curvature perturbation is mixed.
The axions, as usual, contribute only to the
isocurvature perturbation. The resulting total
isocurvature perturbation has mixed
correlation with the adiabatic one.

\par
Most of the parameters of the model but two
were chosen by using criteria of
naturalness and simplicity. We considered two
representative cases for the two remaining
parameters, $\kappa$ and $\lambda$, and
compared the predictions with the first-year
WMAP observations and other CMBR measurements.
We found that, in one of the two cases (model
B), the curvaton and axions contributions to
the CMBR power spectra are important, and that
this leads to a significant disagreement with
the data. In fact, Bayesian model comparison
disfavors this case as compared to the
scale-invariant pure inflaton model with odds
of $10^7$ against 1. The other possibility
which we studied (model A) predicts a
predominantly adiabatic power spectrum from the
inflaton, where the curvaton and axions
contributions are subdominant. We derived upper
bounds for the amplitude of the partial
curvature perturbation from the curvaton in
this case ($A_c\lsim 43.2\times 10^{-5}$ at
95\% c.l.), and noticed that the interplay of
the non-inflaton contributions results in a
later reionization redshift
($z_r=13.3^{+8.1}_{-9.3}$ at 68\% c.l.). Even
though the best-fit likelihood for this case
($-\ln L_* = 721.2$) is better than in the pure
inflaton HZ case, evaluation of the evidence
under the Laplace approximation gives odds of
about 50 to 1 against model A compared to the
pure inflaton HZ case. These odds must be
interpreted with caution, since they do not
take into account the fact that the PQ model
aims at a more fundamental explanatory power
and that it addresses many particle physics
issues which are outside the scope of
inflationary models.

\par
In summary, we have shown that certain regions
of the parameter space ($\kappa\approx 3\times
10^{-2}$ and $10^{-4}\lsim\lambda\lsim 3\times
10^{-4}$) can be excluded on the grounds of
present-day CMBR observations. Quantitative
bounds on the allowed values of $\kappa$ and
$\lambda$ could be derived by treating them as
free parameters in the MC analysis, an
exploration left for future work. Our approach
-- embedding particle physics model building in
the cosmological framework -- pursued the issue
of merging together fundamental theories and
cosmological constraints in a realistic and
viable model for the generation of the
cosmological perturbations. It is encouraging
that modern cosmological observations are now
informative enough as to give useful and robust
guidance along this path.

\section*{ACKNOWLEDGMENTS}
We thank K. Dimopoulos, R. Durrer, S. Leach, D.
Lyth, Ch. Ringeval, M. Sakellariadou and D.
Wands for useful discussions. We are
particularly grateful to K. Dimopoulos and D.
Lyth for communicating to us their results on
the randomization regime for the curvaton field
prior to publication. This work was supported
by the European Union under RTN contracts
HPRN-CT-2000-00148 and HPRN-CT-2000-00152 and
was performed on the Myri cluster owned and
operated by the University of Geneva. R.T. was
partially supported by the European Network
CMBNet and the Swiss National Science
Foundation.

\appendix

\renewcommand{\theequation}
{\thesection\arabic{equation}}
\setcounter{equation}{0}
\renewcommand{\thefigure}
{\thesection\arabic{figure}}
\setcounter{figure}{0}

\newcommand{\gsim}{\,\raise 0.4ex\hbox{$>$}
\kern -0.7em\lower 0.62ex\hbox{$\sim$}\,}

\section{Bayesian inference: a primer}

\subsection{Bayesian parameter estimation}
\label{appendix:params}

\par
Bayesian inference is based on Bayes' theorem,
which is nothing more than rewriting the
definition of conditional probability:
\begin{equation}
\pdf(A\vert B)=\frac{\pdf(B\vert A)\pdf(A)}
{\pdf(B)}
\quad
\text{(Bayes' theorem).}
\end{equation}
In order to clarify the meaning of this
relation, let us write $\params$ (a vector of
$d$ parameters under a model $\mdl$) for
$A$ and $\data$ (the data at hand) for $B$,
obtaining
\begin{eqnarray}
\pdf(\params\vert\data,\mdl)&=&
\frac{\like(\data\vert\params,\mdl)
\pi(\params,\mdl)}{\int_{\Omega}
\like(\data\vert\params,\mdl)
\pi(\params,\mdl)\dr\params}
\nonumber\\
&=&\frac{\like(\data\vert\params,\mdl)
\pi(\params,\mdl)}{\pdf(\data\vert\mdl)},
\label{eq:Bayes_Theorem}
\end{eqnarray}
where $\Omega$ designates the parameter space
(of dimensionality $d$) under model $\mdl$.
This equation relates the {\it posterior
probability} $\pdf(\params\vert\data,\mdl)$ for
the parameters $\params$ of the model $\mdl$
given the data $\data$ to the {\it likelihood
function} $\like(\data\vert\params,\mdl)$ if
the {\it prior probability distribution
function} $\pi(\params,\mdl)$ for the
parameters under the model is known. The latter
is called ``prior'' for short. The quantity in
the denominator is independent of $\params$ and
is called the {\it evidence} of the data for
the model $\mdl$ \cite{MKbook}. The evidence
is the central quantity for model comparison,
as we explain below, but, in the context of
parameter estimation within a model, it is just
an overall multiplicative constant which does
not matter. In short,
\begin{equation}
\text{posterior}=
\frac{\text{likelihood}\times\text{prior}}
{\text{evidence}} .
\end{equation}

\par
The prior distribution contains all the
knowledge about the parameters
before observing the data, i.e. our physical
understanding of the model, our insight into
the experimental setup and its performance,
and in short all our prior scientific
experience. This information is then updated
via Bayes' theorem to the posterior
distribution by multiplying the prior with the
likelihood function which contains the
information coming from the data. The posterior
probability is the base for inference about
$\params$. The most probable value for the
parameters is the one for which the posterior
probability is largest.

\par
Bayes' postulate states that, in the absence of
other arguments, the prior probability should
be assumed to be equal for all values of the
parameters over a certain range
($\params_{\text{min}}\leq\params\leq
\params_{\text{max}}$). This is called a ``flat
prior'', i.e.
\begin{equation}
\pi(\params,\mdl)=\left[ H(\params-
\params_{\text{min}})H(\params_{\text{max}}-
\params)\right]\prod_{j=1}^{d}\frac{1}
{\Delta\theta_j},
\label{eq:flat_prior}
\end{equation}
where $H$ is the Heaviside step function and
$\Delta\theta_j\equiv\theta_{\text{max},j}-
\theta_{\text{min},j}>0$, $\forall \;j$.
Clearly, a flat prior on $\params$ does not
correspond to a flat prior on some other set
$\boldsymbol{\alpha}(\params)$ obtained via a
non-linear transformation, since the two prior
distributions are related via
\begin{equation}
\pi(\params,\mdl)=
\pi(\boldsymbol{\alpha},\mdl)
\frac{\dr\boldsymbol{\alpha}(\params)}
{\dr\params}.
\label{eq:prior}
\end{equation}
A recurrent criticism is that the final
inference depends on the prior which one
chooses to use. However, in a situation in
which the data exhibit a clear preference for a
certain value for a parameter, the posterior is
effectively dominated by the likelihood, and
the choice of prior will not matter much. This
is currently the case, as far as the CMBR is
concerned, for high ``signal to noise''
parameters such as the baryon density.
Constraints on other parameters such as the
curvaton amplitude $A_c$ in model A (upper
band) of our PQ scenario are likely to
depend slightly on the details of the chosen
prior distribution. In other words, constraints
on parameters which are not clearly determined
will suffer from a certain degree of
subjectivity, depending on what prior
$\pi(\params,\mdl)$ we choose on the right
hand side of Eq.~(\ref{eq:Bayes_Theorem}). This
fact should be interpreted as a warning,
telling us that the data are not powerful
enough to clearly single out the parameter
under consideration.

\subsection{Bayes factors}
\label{appendix:factors}

\par
Let us consider two competing models $\mdl_1$
and $\mdl_2$ and ask what is the posterior
probability of each model given the data
$\data$. By Bayes' theorem we have
\begin{equation}
\pdf(\mdl_i\vert\data)\propto
\pdf(\data\vert\mdl_i)\pi(\mdl_i)~~(i=1,2),
\end{equation}
where $\pdf(\data\vert\mdl_i)$ is the
evidence of the data under model $\mdl_i$ and
$\pi(\mdl_i)$ is the prior probability of the
$i$th model before we see the data. The ratio
of the posterior odds for the two competing
models is called {\it Bayes factor}
\cite{Kass}:
\begin{equation}
B_{12}\equiv\frac{\pdf(\mdl_1\vert\data)}
{\pdf(\mdl_2\vert\data)}.
\label{eq:b12}
\end{equation}
Usually, we do not hold any prior beliefs about
the two models and therefore
$\pi(\mdl_1)=\pi(\mdl_2)=1/2$, so that the
Bayes factor reduces to the ratio of the
evidences. The Bayes factor can be interpreted
as an automatic Occam's razor, which disfavors
complex models involving many parameters (see
Ref.~\cite{MKbook} for details) as we discuss
below and demonstrate in the text.

\par
The evidence in favor of $\mdl$ can be
evaluated by performing the integral
\begin{eqnarray}
\pdf(\data\vert\mdl)&=&\int_{\Omega}
\like(\data\vert\params,\mdl)
\pi(\params,\mdl)\dr\params
\nonumber \\
&=&\int_{\Omega}
\bar{\pdf}(\params\vert\data,\mdl)\dr\params,
\end{eqnarray}
where $\bar{\pdf}(\params\vert\data,\mdl)$
designates the non-normalized posterior
probability (i.e. the numerator in the right
hand side of Eq.~(\ref{eq:Bayes_Theorem})).
Computing the above integral from
the MC samples is difficult, since there will
be very few or no samples in the tails of the
likelihood. There are however a number of
approximate methods which can be applied
\cite{DiCiccio}. Most of them rely on the fact
that, for a large number of data points, the
likelihood function will tend to be a
multi-dimensional Gaussian distribution. One
simple approximation is then to expand the
logarithm of the non-normalized posterior to
second order around its peak, which (for flat
prior) occurs at the best-fit parameter choice
$\params_*$, where the likelihood is maximized.
We obtain
\begin{equation}
\ln\frac{\bar{\pdf}(\params\vert\data,\mdl)}
{\bar{\pdf}(\params_*\vert\data,\mdl)}\approx
-\frac{1}{2}(\params-\params_*)^T
{\bf C}^{-1}(\params-\params_*),
\label{eq:laplace}
\end{equation}
where ${\bf C}$ is the covariance matrix
of the model $\mdl$ evaluated at the best-fit
point, which can be estimated from the MC
samples. This is called Laplace approximation
and can be expected to give sensible results if
the non-normalized posterior is reasonably well
described by the multi-dimensional Gaussian
Eq.~(\ref{eq:laplace}). It is then
straightforward to evaluate the evidence by
using the approximate form in
Eq.~(\ref{eq:laplace}) for the non-normalized
posterior, obtaining
\begin{eqnarray}
\pdf(\data\vert\mdl)&=&\int_{\Omega}
\bar{\pdf}(\params\vert\data,\mdl)
\dr\params
\nonumber\\
&\approx&(2\pi)^{\frac{d}{2}}
\bar{\pdf}(\params_*\vert\data,\mdl)
\sqrt{\det{\bf C}}.
\end{eqnarray}
For flat separable priors of the
form in Eq.~(\ref{eq:flat_prior}) we can simply
write, abbreviating ${\bf \Delta}\params
\equiv\prod_{j=1}^{d}\Delta\theta_j$,
\begin{equation}
\bar{\pdf}(\params_*\vert\data,\mdl)=
\like(\data\vert\params_*,\mdl)\frac{1}
{{\bf \Delta}\params},
\label{eq:laplace_2}
\end{equation}
an expression which is still approximately
correct even if we used non-flat priors, and
interpret $\Delta\theta_j$ as the
typical width of the prior pdf (say the
standard deviation along the direction of the
$j$th parameter for a Gaussian distributed
prior). For a Gaussian prior, it is easy to
derive an exact expression analogous to
Eq.~(\ref{eq:laplace_2}), but for simplicity
we will stick to the above form.

\par
For the logarithm of the Bayes factor in the
Laplace approximation, we finally obtain the
following handy expression:
\begin{equation}
\ln B_{12}\approx
\mathcal{L}_{12}+\mathcal{C}_{12}+
\mathcal{F}_{12},
\label{eq:B12}
\end{equation}
where we have defined
\begin{eqnarray}
\mathcal{L}_{12}&\equiv&\ln
\frac{\like(\data\vert\params^{(1)}_*,\mdl_1)}
{\like(\data\vert\params^{(2)}_*,\mdl_2)},
\label{eq:L12}
\\
\mathcal{C}_{12}&\equiv&
\frac{1}{2}\left(\ln \left[(2\pi)^{d^{(1)}-
d^{(2)}}\right]+\ln\frac{\det{\bf C}^{(1)}}
{\det{\bf C}^{(2)}}\right),
\label{eq:C12}
\\
\mathcal{F}_{12}&\equiv&
\ln\frac{{\bf \Delta}\params^{(2)}}
{{\bf \Delta}\params^{(1)}}
\label{eq:F12},
\end{eqnarray}
where quantities referring to the model
$\mdl_i$ ($i=1,2$) are indicated by a
superscript $(i)$. The term
$\mathcal{L}_{12}$ is the logarithm of the
ratio of the best-fit likelihoods. From a
frequentist point of view, this quantity is
approximately $\chi^2$ distributed, and thus
it is common practice to apply to it a
goodness-of-fit test to assess whether the
extra parameters have produced a
``significant'' increase of the quality of
fit. If the model $\mdl_1$ contains more
parameters than the model $\mdl_2$, then
$\mdl_1$ should show an improved fit to the
data, i.e. we should have $\mathcal{L}_{12}>0$,
unless the extra parameters are useless, in
which case $\mathcal{L}_{12}=0$. In any case,
a goodness-of-fit test alone does not say
anything about the structure of the parameter
space for the model under consideration, since
it is limited to the maximum likelihood point.
But Bayesian evidence does contain two further
pieces of information, $\mathcal{C}_{12}$ and
$\mathcal{F}_{12}$, which taken together are
sometimes referred to as ``Occam's factor''.
Here we prefer to consider these terms
separately to help distinguishing their
different origin. The term $\mathcal{C}_{12}$
describes the structure of the posterior shape
in the Gaussian approximation. Since the
determinant is the product of the eigenvalues
of the covariance matrix, which represent the
standard deviations squared along the
corresponding eigenvectors in the parameter
space of the model, it follows that if $\mdl_1$
has a narrower posterior than $\mdl_2$, then
$\mathcal{C}_{12}<0$, thereby disfavoring
$\mdl_1$. This apparent contradiction (how can
a model with smaller errors display a smaller
evidence?) is resolved when we take into
account the term $\mathcal{F}_{12}$, which
describes the prior available parameter space
under each model. The sum of the terms
$\mathcal{C}_{12}$ and $\mathcal{F}_{12}$ thus
disfavors the model with the largest volume of
``wasted'' parameter space when the data
arrive. A more complex model $\mdl_1$ -- having
a large number of parameters and thus a large
volume of prior accessible parameter space --
will naturally fit the data better due to its
flexibility, i.e. we will have
$\mathcal{L}_{12}>0$, but it will be penalized
for introducing extra dimensions in parameter
space, i.e. the sum $\mathcal{C}_{12}+
\mathcal{F}_{12}$ will be negative. In summary,
the Bayes factor tends to select the model
which exhibits an optimal trade-off between
simplicity and quality of fit.

\par
It should be clear that the choice of priors
plays an important role in Bayesian model
comparison (testing) via its impact on the term
$\mathcal{F}_{12}$. In particular, prior pdf's
used in evaluating the Bayes factor must be
proper, i.e. normalizable, so that we can
impose the normalization condition
\begin{equation}
\int_{\Omega}\pi(\params,\mdl)\dr\params=1.
\end{equation}
Although generally a strong dependence on the
choice of priors is regarded as suspicious, in
this case, we should consider the role of
$\mathcal{F}_{12}$ as a way to implement
{\it a priori} model features into the Bayes
factor, as we show in the text of the paper.

\def\ijmp#1#2#3{{Int. Jour. Mod. Phys.}
{\bf #1},~#3~(#2)}
\def\plb#1#2#3{{Phys. Lett. B }{\bf #1},~#3~(#2)}
\def\zpc#1#2#3{{Z. Phys. C }{\bf #1},~#3~(#2)}
\def\prl#1#2#3{{Phys. Rev. Lett.}
{\bf #1},~#3~(#2)}
\def\rmp#1#2#3{{Rev. Mod. Phys.}
{\bf #1},~#3~(#2)}
\def\prep#1#2#3{{Phys. Rep. }{\bf #1},~#3~(#2)}
\def\prd#1#2#3{{Phys. Rev. D }{\bf #1},~#3~(#2)}
\def\npb#1#2#3{{Nucl. Phys. }{\bf B#1},~#3~(#2)}
\def\npps#1#2#3{{Nucl. Phys. B (Proc. Sup.)}
{\bf #1},~#3~(#2)}
\def\mpl#1#2#3{{Mod. Phys. Lett.}
{\bf #1},~#3~(#2)}
\def\arnps#1#2#3{{Annu. Rev. Nucl. Part. Sci.}
{\bf #1},~#3~(#2)}
\def\sjnp#1#2#3{{Sov. J. Nucl. Phys.}
{\bf #1},~#3~(#2)}
\def\jetp#1#2#3{{JETP Lett. }{\bf #1},~#3~(#2)}
\def\app#1#2#3{{Acta Phys. Polon.}
{\bf #1},~#3~(#2)}
\def\rnc#1#2#3{{Riv. Nuovo Cim.}
{\bf #1},~#3~(#2)}
\def\ap#1#2#3{{Ann. Phys. }{\bf #1},~#3~(#2)}
\def\ptp#1#2#3{{Prog. Theor. Phys.}
{\bf #1},~#3~(#2)}
\def\apjl#1#2#3{{Astrophys. J. Lett.}
{\bf #1},~#3~(#2)}
\def\n#1#2#3{{Nature }{\bf #1},~#3~(#2)}
\def\apj#1#2#3{{Astrophys. J.}
{\bf #1},~#3~(#2)}
\def\anj#1#2#3{{Astron. J. }{\bf #1},~#3~(#2)}
\def\apjs#1#2#3{{Astrophys. J. Suppl.}
{\bf #1},~#3~(#2)}
\def\mnras#1#2#3{{Mon. Not. Roy. Astron. Soc.}
{\bf #1},~#3~(#2)}
\def\grg#1#2#3{{Gen. Rel. Grav.}
{\bf #1},~#3~(#2)}
\def\s#1#2#3{{Science }{\bf #1},~#3~(#2)}
\def\baas#1#2#3{{Bull. Am. Astron. Soc.}
{\bf #1},~#3~(#2)}
\def\ibid#1#2#3{{\it ibid. }{\bf #1},~#3~(#2)}
\def\cpc#1#2#3{{Comput. Phys. Commun.}
{\bf #1},~#3~(#2)}
\def\astp#1#2#3{{Astropart. Phys.}
{\bf #1},~#3~(#2)}
\def\epjc#1#2#3{{Eur. Phys. J. C}
{\bf #1},~#3~(#2)}
\def\nima#1#2#3{{Nucl. Instrum. Meth. A}
{\bf #1},~#3~(#2)}
\def\jhep#1#2#3{{J. High Energy Phys.}
{\bf #1},~#3~(#2)}
\def\lnp#1#2#3{{Lect. Notes Phys.}
{\bf #1},~#3~(#2)}
\def\appb#1#2#3{{Acta Phys. Polon. B}
{\bf #1},~#3~(#2)}
\def\fcp#1#2#3{{Fund. Cos. Phys.}
{\bf #1},~#3~(#2)}
\def\ptps#1#2#3{{Prog. Theor. Phys. Suppl.}
{\bf #1},~#3~(#2)}
\def\ss#1#2#3{{Statist. Sci.}
{\bf #1},~#3~(#2)}
\def\jasa#1#2#3{{J. Amer. Statist. Assoc.}
{\bf #1},~#3~(#2)}
\def\jcap#1#2#3{{J. Cosmol. Astropart. Phys.}
{\bf #1},~#3~(#2)}
\def\bm#1#2#3{{Biometrika}
{\bf #1},~#3~(#2)}
\def\ams#1#2#3{{Ann. Math. Stat.}
{\bf #1},~#3~(#2)}

\end{document}